\documentclass[preprint,preprintnumbers,jcp,showpacs,superscriptaddress]{revtex4-1}
\usepackage{graphicx}
\usepackage{dcolumn}
\usepackage[latin1]{inputenc}
\usepackage{latexsym}
\usepackage{amscd,amssymb,amsthm}
\usepackage{amsmath}
\usepackage{amsfonts}
\usepackage{color}
\usepackage{changes}

\DeclareMathAlphabet\mathbfcal{OMS}{cmsy}{b}{n}
\def\bA{{\bf A}}

\def\dulR{{\underline{\underline{\bf R}}}}
\def\dulr{{\underline{\underline{\bf r}}}}

\begin{document}

\title{Quantum-Classical Non-Adiabatic Dynamics: Coupled- vs. Independent-Trajectory Methods}

\author{Federica Agostini}
\affiliation{Max-Planck Institut f\"ur Mikrostrukturphysik, Weinberg 2, D-06120 Halle, Germany}
\author{Seung Kyu Min}
\affiliation{Department of Chemistry, School of Natural Science, Ulsan 
National Institute of Science and Technology (UNIST), 50 UNIST-gil, Ulsan 
44919, Republic of Korea}
\email{skmin@unist.ac.kr}
\author{Ali Abedi}
\affiliation{Nano-Bio Spectroscopy group and European Theoretical Spectroscopy Facility (ETSF), Dpto. F{\'i}sica de Materiales, 
Universidad del Pa{\'i}s Vasco, Centro de F{\'i}sica de Materiales 
CSIC-UPV/EHU-MPC and DIPC, Av. Tolosa 72, E-20018 San Sebasti{\'a}n, Spain}
\author{E. K. U. Gross}
\affiliation{Max-Planck Institut f\"ur Mikrostrukturphysik, Weinberg 2, D-06120 Halle, Germany}

%\date{\today}
\begin{abstract}
Trajectory-based mixed quantum-classical approaches to coupled electron-nuclear dynamics suffer from well-studied problems such as the lack of (or incorrect account for) decoherence in the trajectory surface hopping method and the inability of reproducing the spatial splitting of a nuclear wave packet in Ehrenfest-like dynamics. In the context of electronic non-adiabatic processes, 
these problems can result in wrong predictions for quantum populations and in 
unphysical outcomes for the nuclear dynamics. In this paper we propose a 
solution to these issues by approximating the coupled electronic and nuclear 
equations within the framework of the exact factorization of the 
electron-nuclear wave function. We present a simple quantum-classical scheme
based on coupled classical trajectories, and test it against 
the full quantum mechanical solution from wave packet dynamics for some model 
situations which represent particularly challenging 
problems for the above-mentioned traditional methods.
\end{abstract}

\maketitle

\section{Introduction}
Non-adiabatic effects often play an important role in the coupled dynamics 
of electrons and nuclei. 
Typical examples of processes whose description requires to 
explicitly account for non-adiabatic electronic transitions induced by the 
nuclear motion are vision~\cite{cerulloN2010, schultenBJ2009, ishidaJPCB2012}, 
photo-synthesis~\cite{tapaviczaPCCP2011, flemingN2005}, 
photo-voltaics~\cite{rozziNC2013, silvaNM2013, jailaubekovNM2013}, and charge 
transport through molecular junctions.~\cite{meyer_NP2010, wegewijs_PRL2010, 
thoss_PRL2011, todorov_NN2009} In all these cases, the electronic effect on 
the nuclei cannot be expressed by a single adiabatic potential energy surface 
(PES), corresponding to the occupied eigenstate of the Born-Oppenheimer (BO) Hamiltonian. The exact numerical treatment would in fact require the inclusion of several adiabatic 
PESs that are coupled via electronic non-adiabatic transitions in regions of 
strong coupling, such as avoided crossings or conical intersections. Based on this 
theoretical picture, numerical methods that retain a quantum description of 
the nuclei have been successfully employed in many applications, e.g. 
multiple-spawning~\cite{martinezCPL1996, martinezJPC1996, martinezACR2006},  
multiconfiguration time-dependent {Hartree}~\cite{cederbaumCPL1990, 
burghardtJCP1999, thossCM2004}, or non-adiabatic Bohmian 
dynamics~\cite{tavernelliPCCP2011, tavernelliPRA2013, Albareda_PRL2014, Albareda_JPCL2015, wyattJCP2001, wyattJCP2002, parlantJPCA2007}, but actual calculations 
become unfeasible for systems comprising hundreds or thousands of atoms. 
Promising alternatives are therefore those approaches that involve a classical or 
quasi-classical treatment of nuclear motion coupled
non-adiabatically to the quantum mechanical motion of the
electrons~\cite{pechukasPR1969_1, pechukasPR1969_2, ehrenfest, 
shalashilin_JCP2009, tully1971, tully1990, tully1998, tullyJCP2000, prezhdoPRL2001, shenvi-subotnikJCP2011, landryJCP2013_2, kapral-ciccotti, kapralARPC2006, io, gevaJCP2004, schuetteJCP2001, martensJCP1997, bonellaCP2001, cokerJCP2012, millerJCP1997, thossPRL1997, millerJCP2007, millerJPCA2009, thossARPC2004, hermannARPC1994, hermanJCP2005, truhlarFD2004, truhlarACR2006, 
burghardtJCP2011, jangJCP2012, vanicekJCP2012, takatsukaPRA2010, 
bonellaJCP2005, marxPRL2002, Gross_EPL2014, prezhdo, tannorJCP2012}. There are two main 
challenges that theory has to face in this context: (i) The splitting of the 
nuclear wave packet induced by non-adiabatic couplings needs to be captured in 
the trajectory-based description.  
(ii) This requires a clear-cut definition of the classical force in situations 
when more than a single adiabatic PES is involved.

In the present paper, we attack these two points within the framework of the 
exact factorization of the electron-nuclear wave function~\cite{Gross_PRL2010, 
Gross_JCP2012}. When the solution of the time-dependent Schr\"odinger equation 
(TDSE) is written as a single product of a nuclear wave function and an 
electronic factor, which parametrically depends on the nuclear configuration, 
two coupled equations of motion for the two components of the full wave 
function are derived from the TDSE. These equations contain the answer to the 
above questions. The purpose of this paper is to provide a systematic procedure to 
develop the necessary approximations when the nuclei are represented in terms 
of classical trajectories. The final result will be a mixed quantum-classical (MQC) 
algorithm~\cite{Gross_PRL2015} that we have implemented and tested in various 
non-adiabatic situations. The numerical scheme presented here is based on the 
analysis performed so far in the framework of the exact factorization. In 
previous work, we have first focused on the nuclear dynamics, by analysing the 
time-dependent potentials~\cite{Gross_PRL2013} of the theory, with particular 
attention devoted to understanding the fundamental properties that need to be 
accounted for when introducing approximations. Moreover, we have analyzed the 
suitability of the classical and quasi-classical treatment~\cite{Gross_MP2013, 
Gross_JCP2015, Agostini_ADP2015} of nuclear dynamics, and we have proposed an 
independent-trajectory MQC scheme~\cite{Gross_EPL2014, Gross_JCP2014} to solve 
the coupled electronic and nuclear equations within the factorization framework approximately. This result can be viewed as the lowest order version 
of the algorithm presented here, where refined and more accurate 
approximations have been introduced.

The new algorithm\cite{Gross_PRL2015}, which is based on a coupled-trajectory 
(CT) description of nuclear dynamics, will be presented (i) paying particular 
attention to the physical justification of the approximations introduced, (ii) 
describing in detail the fundamental equations of the CT-MQC scheme in the 
form they are implemented in practice, (iii) supporting the analytical derivation 
with abundant numerical proof of the suitability of the procedure for 
simulating non-adiabatic dynamics. In particular, we show how the CT procedure 
compares with the previous MQC~\cite{Gross_EPL2014, Gross_JCP2014} algorithm 
based on the exact factorization and with the Ehrenfest~\cite{ehrenfest} and 
trajectory surface hopping (TSH)~\cite{tully1990} approaches. This comparison 
will demonstrate that the CT scheme is able (a) to properly describe the splitting 
of a nuclear wave packet after the passage through an avoided crossing, 
related to the steps~\cite{Gross_PRL2013} in the time-dependent potential of 
the theory, overcoming the limitations of MQC and Ehrenfest, (b) to correctly 
account for electronic decoherence effects, thus proposing a solution to the 
over-coherence problem~\cite{truhlarACR2006, shenvi-subotnikJCP2011, 
shenviJCP2012, subotnikJCP2011_1, shenviJCP2011_2, subotnikJPCA2011, 
prezhdoJCP2014, landryJCP2013_2, tretiakJCP2013, prezhdoJCP1999, 
rosskyJCP1995, rosskyJCP1997, truhlarJCP2008, schwartzJCP2005, persicoJCP2007, 
persicoJCP2010, persicoJCP2014} affecting TSH.

The paper is organized as follows. The first section focusses on the theory. We construct the algorithm, describe the used approximations and derive the equations that have been implemented. The following section gives a schematic overview of the algorithm, summarizing the results of the first section. The steps to be implemented are described here. The third section shows the numerical results for one-dimensional two-state model systems, representing situations of (a) single avoided crossing, (b) dual avoided crossing, (c) extended coupling with reflection and (d) double arch. Conclusions are drawn in the fourth section.

\section{Coupled-trajectory mixed quantum-classical algorithm}
The exact factorization of the electron-nuclear wave function~\cite{Gross_PRL2010, Gross_JCP2012} provides the theoretical background for the development of the algorithm described and tested in this paper. Since the theory has been extensively presented in previous work~\cite{Gross_PRL2013, Gross_MP2013, Gross_EPL2014, Gross_JCP2014, Gross_JCP2015, Agostini_ADP2015, Gross_PRL2015, Gross_PTRSA2014, Min_PRL2014, Requist_PRA2015, Suzuki_PRA2014, Suzuki_PCCP2015, Khosravi_PRL2015, Schild_JPCA2016}, we only refer to Section SI.1 of the Supporting Information for a comprehensive review of the theory. 

It has been proved~\cite{Gross_PRL2010, Gross_JCP2012} that the solution of the time-dependent Schr\"odinger equation (TDSE) $[\hat T_n+\hat H_{BO}]\Psi = i\hbar\partial_t\Psi$ of a combined system of electrons and nuclei can be written as the product: $\Psi(\dulr,\dulR,t)=\Phi_\dulR(\dulr,t)\chi(\dulR,t)$. Here, $\chi(\dulR,t)$ is the nuclear wave function, which yields the exact nuclear many-body density and current density, whereas $\Phi_\dulR(\dulr,t)$ is the electronic conditional wave function, which parametrically depends on the nuclear configuration $\dulR$. The squared modulus of $\Phi_\dulR(\dulr,t)$ is normalized to unity $\forall\,\dulR,t$, thus $|\Phi_\dulR(\dulr,t)|^2$ can be interpreted as a conditional probability. The equations of motion describing the time evolution of the two terms of the product are derived by determining the stationary variations~\cite{frenkel, alonsoJCP2013, Gross_JCP2013} of the quantum mechanical action with respect to $\Phi_\dulR(\dulr,t)$ and $\chi(\dulR,t)$, yielding
\begin{align}
&\Bigg(\hat H_{BO}+\hat U_{en}\left[\Phi_\dulR,\chi\right]-\epsilon(\dulR,t)\Bigg)\Phi_{\dulR}(\dulr,t)=i\hbar\partial_t \Phi_{\dulR}(\dulr,t)\label{eqn: exact electronic eqn}\\
&\Bigg(\sum_{\nu=1}^{N_n}\frac{\left[-i\hbar\nabla_\nu+\bA_\nu(\dulR,t)\right]^2}{2M_\nu} + \epsilon(\dulR,t)\Bigg)\chi(\dulR,t)=i\hbar\partial_t \chi(\dulR,t). \label{eqn: exact nuclear eqn}
\end{align}
Here the electron-nuclear coupling operator (ENCO) is defined as
\begin{align}\label{eqn: enco}
\hat U_{en}&\left[\Phi_\dulR,\chi\right]=\nonumber\\
&\sum_{\nu=1}^{N_n}\frac{1}{M_\nu}\Bigg[\frac{\left[-i\hbar\nabla_\nu-\bA_\nu(\dulR,t)\right]^2}{2}+\left(\frac{-i\hbar\nabla_\nu\chi}{\chi}+\bA_\nu(\dulR,t)\right)\Big(-i\hbar\nabla_\nu-\bA_{\nu}(\dulR,t)\Big)\Bigg].
\end{align}
The time-dependent vector potential (TDVP) and time-dependent potential energy surface (TDPES) are given by the expressions
\begin{align}
 \bA_{\nu}\left(\dulR,t\right) &= \left\langle\Phi_\dulR(t)\right|-i\hbar\nabla_\nu\left.\Phi_\dulR(t)
 \right\rangle_\dulr\,\label{eqn: tdvp} \\
 \epsilon(\dulR,t)&=\left\langle\Phi_\dulR(t)\right|\hat{H}_{BO}+\hat U_{en}^{coup}-i\hbar\partial_t\left|
 \Phi_\dulR(t)\right\rangle_\dulr, \label{eqn: tdpes}
\end{align}
respectively. The numerical procedure referred to as coupled-trajectory mixed quantum-classical (CT-MQC) algorithm, introduced in previous work~\cite{Gross_PRL2015} to solve Eqs.~(\ref{eqn: exact electronic eqn}) and~(\ref{eqn: exact nuclear eqn}), will be presented here paying special attention to justifying the approximations and describing the implementation of the algorithm.

\subsection{The Lagrangian frame}
\begin{figure}[h!]
 \begin{center}
  \includegraphics[width=0.8\textwidth]{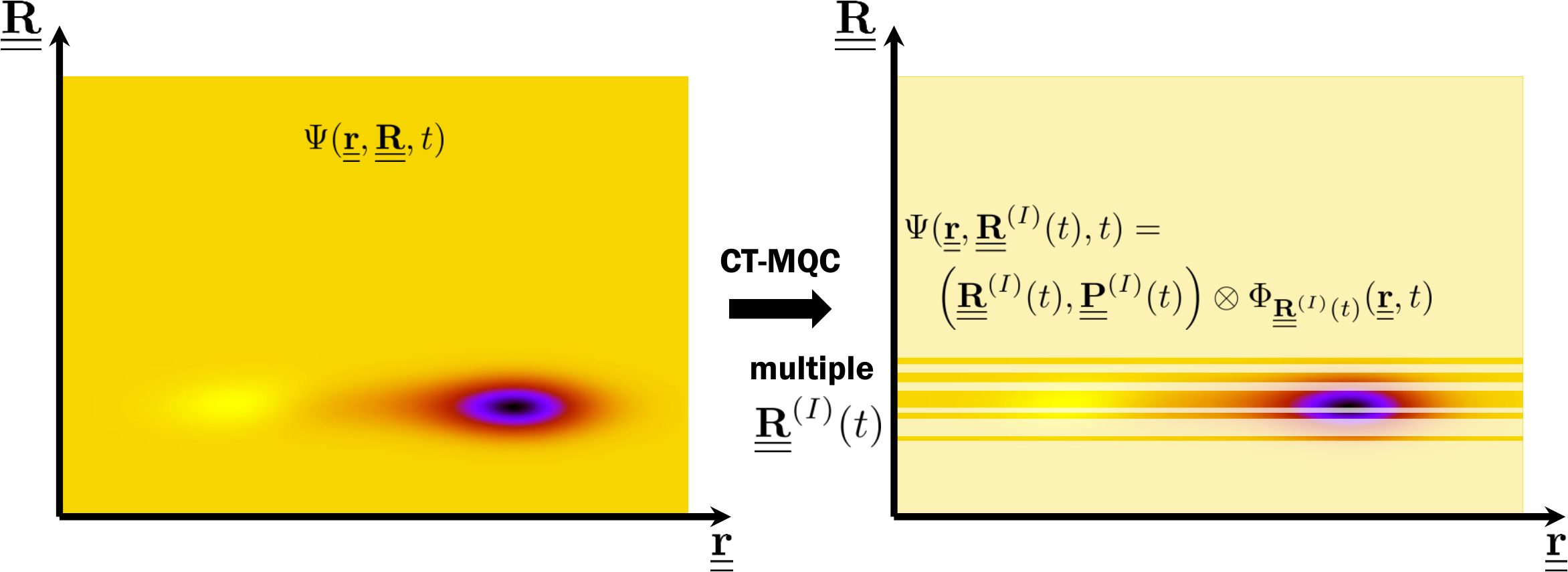}
 \end{center}
 \caption{Full molecular wave function (left) as function of electronic ($\dulr$) and nuclear ($\dulR$) coordinates and its approximation (right) on a set of nuclear trajectories ($\dulR^{(I)}(t)$'s). In this case, the molecular wave function is known only at the given positions $\dulR^{(I)}(t)$.}
 \label{fig: scheme}
\end{figure}

Equations~(\ref{eqn: exact electronic eqn}) and~(\ref{eqn: exact nuclear eqn}) 
can be represented on a fixed spatial grid and propagated in time, to compute 
the evolution of the electronic and nuclear wave functions. However, our aim 
is to represent the dynamics of quantum nuclei via the motion of a set of 
coupled classical trajectories, $\dulR^{(I)}(t)$, whose behavior can be 
assimilated to that of a moving grid. In turn, the dynamics of the electrons 
is governed by Eq.~(\ref{eqn: exact electronic eqn}) along each nuclear 
trajectory. Figure~\ref{fig: scheme} shows the idea behind such representation 
of the nuclei in terms of classical trajectories: on the left, a molecular 
wave function (or better the density from a molecular wave function) is 
plotted on a fixed grid in $\dulr,\dulR$ space; on the right, some information 
along the $\dulR$-direction is lost, since the wave function is available only 
at the positions of the trajectories, located where the nuclear density is larger. 
Therefore, it seems more natural to work in a reference frame that moves with 
the trajectories, the Lagrangian frame, rather than fixed Eulerian frame, and 
to introduce the approximations starting from this picture. In the Lagrangian 
frame time-derivatives are calculated ``along the flow'', thus all partial 
time-derivatives have to be replaced by total derivatives, using the chain 
rule $d/dt = \partial_t + \sum_\nu \mathbf V_\nu\cdot \nabla_\nu$. Here, the 
quantity $\mathbf V_\nu$ is the velocity of the moving grid point, i.e. the velocity of each trajectory which can be determined from the equations below.

We introduce now the \textit{first approximation}, namely we derive the classical Newton's equation from the TDSE~(\ref{eqn: exact nuclear eqn}). We have previously proposed~\cite{Gross_JCP2014, Gross_EPL2014, Gross_PRL2015} a derivation based on the complex phase-representation~\cite{van-vleck} of $\chi(\dulR,t)$. Here, we present an alternative procedure~\cite{Gross_JCP2015}, by writing the nuclear wave function in polar form, $\chi(\dulR,t)=|\chi(\dulR,t)|e^{(i/\hbar)S(\dulR,t)}$. Then, the real part of Eq.~(\ref{eqn: exact nuclear eqn}) yields
\begin{align}
\frac{\partial}{\partial t} S(\dulR,t) = 
-\sum_{\nu=1}^{N_n}\frac{\left[\nabla_\nu S(\dulR,t)+{\bf 
A}_\nu(\dulR,t)\right]^2}{2M_\nu}-\epsilon(\dulR,t)+\hbar^2\sum_{\nu=1}^{N_n}\frac{1}{2M_\nu}\frac{\nabla_\nu^2|\chi(\dulR,t)|}{|\chi(\dulR,t)|},\label{eqn: QHJE}
\end{align}
a Hamilton-Jacobi equation (HJE) in the presence of the TDVP $\bA_\nu(\dulR,t)$ and of a potential term (last term in Eq.~(\ref{eqn: QHJE})) known in the framework of Bohmian dynamics as quantum potential. The imaginary part of Eq.~(\ref{eqn: exact nuclear eqn}) yields a continuity equation for the nuclear density. Neglecting the quantum potential, Eq.~(\ref{eqn: QHJE}) becomes a (standard) classical HJE,
\begin{equation}\label{eqn: HJE}
\dot S(\dulR,t) = -\sum_{\nu=1}^{N_n}\Bigg[\frac{\left[\nabla_\nu S(\dulR,t)+{\bf A}_\nu(\dulR,t)\right]^2}{2M_\nu}-\mathbf V_\nu\cdot \nabla_\nu S(\dulR,t)\Bigg]-\epsilon(\dulR,t),
\end{equation}
where $\dot S$ stands for the full time derivative of $S$. In previous work~\cite{Gross_JCP2014, Gross_EPL2014}, rather than $S(\dulR,t)$, the HJE contained $S_0(\dulR,t)$, the lowest order term in an expansion of the complex phase in powers of Planck's constant. The comparison with this result~\cite{Agostini_ADP2015} allows us to define the canonical momentum of the moving grid $M_\nu\mathbf V_\nu = \nabla_\nu S_0(\dulR,t)+{\bf A}_\nu(\dulR,t)=\mathbf P_\nu$. Taking the spatial derivative $\nabla_{\nu'}$ on both sides, Eq.~(\ref{eqn: HJE}) reduces to a classical evolution equation for the 
characteristics
\begin{align}\label{eqn: force before gauge}
\dot{\mathbf P}_\nu(t)\Big|_{\dulR^{(I)}(t)}=&-\nabla_{\nu}\left(\epsilon(\dulR,t)+\sum_{\nu'=1}^{N_n}{\bf A}_{\nu'}(\dulR,t)\cdot \frac{\bf P_{\nu'}(\dulR,t)}{M_\nu}\right)+\dot{\bf A}_\nu(\dulR,t)\Bigg|_{\dulR^{(I)}(t)}.
\end{align}
At time $t$ all quantities are evaluated at the grid point $\dulR^{(I)}(t)$.

Eq.~(\ref{eqn: force before gauge}) identifies the classical force used to evolve the grid points $\dulR^{(I)}(t)$. In the proposed implementation, (i) at the initial time $N_{traj}$ points will be sampled from the probability distribution associated to the initial nuclear density $|\chi(\dulR,t=0)|^2$, (ii) the time-evolved of those points will give information about the molecular wave function in the regions of nuclear configuration space of large probability at every time, as shown in Fig.~\ref{fig: scheme} (iii) the nuclear density at time $t$ can be reconstructed from a histogram~\cite{Gross_EPL2014, Gross_JCP2014, Gross_JCP2015} representing the distribution of classical trajectories.

\subsection{Choice of the gauge}
It is straightforward to prove, as described in Section~SI.1 of the Supporting Information, that the product form the electron-nuclear wave function is invariant under a $(\dulR,t)$-phase transformation of $\Phi_\dulR(\dulr,t)$ and $\chi(\dulR,t)$. In order to fix this gauge freedom, Eqs.~(\ref{eqn: exact electronic eqn}) and~(\ref{eqn: exact nuclear eqn}) have to be solved with an additional constraint, which is chosen in this case to be
\begin{align}\label{eqn: gauge}
\epsilon(\dulR,t)+\sum_{\nu=1}^{N_n}{\bf A}_{\nu}(\dulR,t)\cdot \frac{{\bf P}_{\nu}(\dulR,t)}{M_\nu}=0.
\end{align}
This condition is imposed in both the electronic and nuclear equations. In 
particular, Eq.~(\ref{eqn: force before gauge}) simplifies to 
$\dot{\mathbf P}_\nu(t)=\dot{\mathbf A}_\nu(t)$. When the 
electronic~(\ref{eqn: exact electronic eqn}) and nuclear~(\ref{eqn: exact 
nuclear eqn}) equations are integrated by imposing Eq.~(\ref{eqn: gauge}), 
this gauge condition is automatically satisfied but it should be checked 
during the evolution in order to test the numerical accuracy.  

\subsection{Time-dependent potential energy surface}
The expression of the TDPES in the Lagrangian frame becomes
\begin{align}\label{eqn: approx TDPES}
\epsilon^{apx}(\dulR,t)=\left\langle\Phi_\dulR(t)\right|\hat{H}_{BO}\left| \Phi_\dulR(t)\right\rangle_\dulr-i\hbar\left\langle\Phi_\dulR(t)\right|\left. \dot{\Phi}_\dulR(t)\right\rangle_\dulr-\sum_{\nu=1}^{N_n}\frac{\bf P_\nu}{M_\nu}\cdot\mathbf A_\nu(\dulR,t),
\end{align}
where we have introduced here the \textit{second approximation}, namely we have neglected the first term on the right-hand-side in the definition~(\ref{eqn: enco}) of the ENCO. This term contains second-order derivatives of the electronic wave function with-respect-to the nuclear positions, whose calculation requires a larger numerical effort than first-order derivatives. Studies have shown that this term is indeed negligible~\cite{Scherrer_JCP2015} if compared to the leading contribution, i.e. the second term on the right-hand-side of Eq.~(\ref{eqn: enco}). Furthermore, the term containing the second-order derivatives is the only term that contributes to the TDPES, the other being identically zero when averaged over $\Phi_\dulR(\dulr,t)$. Therefore, in order to maintain gauge-invariance, the ENCO cannot appear~\cite{Gross_PRL2013, Gross_MP2013} in the expression of the TDPES, as it is clear in Eq.~(\ref{eqn: approx TDPES}). Moreover, Eq.~(\ref{eqn: approx TDPES}) is obtained by replacing the partial time-derivative from Eq.~(\ref{eqn: tdpes}) with the total time-derivative. The additional term from the chain rule used above contains $\nabla_\nu$, which leads to the appearance of the TDVP when averaged over the electronic wave function. 

\subsection{The Born-Huang expansion}
Classical trajectories evolve according to the force in Eq.~(\ref{eqn: force before gauge}), where the gauge condition~(\ref{eqn: gauge}) is imposed at each time, and are coupled to an approximated form of the electronic equation, namely
\begin{align}
i\hbar\dot{\Phi}_\dulR(\dulr,t) =& \Bigg[\hat H_{BO}+\sum_{\nu=1}^{N_n}\frac{1}{M_\nu}\left(\frac{-i\hbar\nabla_\nu\chi(\dulR,t)}{\chi(\dulR,t)}+{\bf A}_\nu(\dulR,t)-{\mathbf P}_\nu(t)\right)\left(-i\hbar\nabla_\nu\right)\Bigg]\Phi_\dulR(\dulr,t)\nonumber \\
&-\Bigg[\sum_{\nu=1}^{N_n}\frac{1}{M_\nu}\left(\frac{-i\hbar\nabla_\nu\chi(\dulR,t)}{\chi(\dulR,t)}+{\bf A}_\nu(\dulR,t)\right){\bf A}_\nu(\dulR,t)+\epsilon^{apx}(\dulR,t)\Bigg]\Phi_\dulR(\dulr,t),\label{eqn: apx electronic eqn 1}
\end{align}
where we have replaced once again the partial time-derivative with the total derivative, according to
$ \partial_t \Phi_{\dulR}(\dulr,t)\big|_{\dulR^{(I)}(t)} =\dot{\Phi}_{\dulR}(\dulr,t) - \sum_{\nu=1}^{N_n} \frac{\mathbf P_\nu(t)}{M_\nu}\cdot \nabla_\nu \Phi_{\dulR}(\dulr,t)\big|_{\dulR^{(I)}(t)}$.
Notice that all quantities in Eq.~(\ref{eqn: apx electronic eqn 1}) are evaluated along the classical trajectories $\dulR^{(I)}(t)$ and that no further approximation than those described so far has been used to derive this expression. In the numerical scheme proposed here~\cite{Gross_PRL2015}, the electronic wave function is expanded according to the so-called Born-Huang expansion in terms of the adiabatic states $\varphi_\dulR^{(l)}(\dulr)$, which are eigenstates of the Hamiltonian $\hat H_{BO}$ with eigenvalues $\epsilon_{BO}^{(l)}(\dulR)$. Inserting the expansion
\begin{align}\label{eqn: BO expansion}
\Phi_{\dulR}(\dulr,t) = \sum_l C_l\left(\dulR,t\right)\varphi_\dulR^{(l)}(\dulr)
\end{align}
in Eq.~(\ref{eqn: apx electronic eqn 1}), we derive a set of coupled partial differential equations for the coefficients $C_l\left(\dulR,t\right)$
\begin{align}\label{eqn: apx electronic eqn before gauge}
\dot C_l^{(I)}(t) = \frac{-i}{\hbar}\Bigg[\epsilon_{BO}^{(l)(I)}-{\epsilon^{apx}}^{(I)}(t)-\sum_{\nu=1}^{N_n}\frac{\mathbf P_\nu^{(I)}(t)+i\boldsymbol{\mathcal P}_\nu^{(I)}(t)}{M_\nu}\cdot\mathbf A_\nu(t)\Bigg]C_l^{(I)}(t)-&\sum_{\nu=1}^{N_n}\frac{i\boldsymbol{\mathcal P}_\nu^{(I)}(t)}{M_\nu}\nabla_\nu C_l^{(I)}(t)\nonumber\\
-\sum_{\nu=1}^{N_n}\frac{\mathbf P_\nu^{(I)}(t)+i\boldsymbol{\mathcal P}_\nu^{(I)}(t)}{M_\nu}\sum_k &C_k^{(I)}(t)\mathbf d_{\nu,lk}^{(I)}.
\end{align}
The non-adiabatic coupling vectors (NACVs) have been introduced in the above expression, i.e. $\mathbf d_{\nu,lk}^{(I)}=\langle\varphi_{\dulR^{(I)}(t)}^{(l)}|\nabla_\nu\varphi_{\dulR^{(I)}(t)}^{(k)}\rangle_\dulr$. We have used here a superscript $(I)$ to indicate that all quantities depending on $\dulR$ are to be calculated at the position $\dulR^{(I)}(t)$ at time $t$. Henceforth, the spatial dependence will only be indicated by such superscript symbol. In deriving Eq.~(\ref{eqn: apx electronic eqn before gauge}), we have employed the polar representation of the nuclear wave function in $\hat U_{en}$ of Eq.~(\ref{eqn: enco}), i.e.
\begin{align}
\frac{-i\hbar\nabla_\nu\chi^{(I)}(t)}{\chi^{(I)}(t)}+\mathbf A_\nu^{(I)}(t) &= \left[\nabla_\nu S^{(I)}(t)+\mathbf A_\nu^{(I)}(t)\right] + i\frac{-\hbar\nabla_\nu\left|\chi^{(I)}(t)\right|}{\left|\chi^{(I)}(t)\right|}\nonumber \\
&=\mathbf P_\nu^{(I)}(t) + i\boldsymbol{\mathcal P}_\nu^{(I)}(t).\label{eqn: quantum momentum 1}
\end{align}
This equation contains the quantities $\mathbf P_\nu^{(I)}(t)$ and $\boldsymbol{\mathcal P}_\nu^{(I)}(t)$, which will be discussed in detail below.

Let us now introduce the gauge condition~(\ref{eqn: gauge}) in Eq.~(\ref{eqn: apx electronic eqn before gauge}), namely
\begin{align}\label{eqn: apx electronic eqn}
\dot C_l^{(I)}(t) = \frac{-i}{\hbar}\Bigg[\epsilon_{BO}^{(l)(I)}-\sum_{\nu=1}^{N_n}\frac{i\boldsymbol{\mathcal P}_\nu^{(I)}(t)}{M_\nu}\cdot\mathbf A_\nu(t)\Bigg]C_l^{(I)}(t)-\sum_{\nu=1}^{N_n}\frac{i\boldsymbol{\mathcal P}_\nu^{(I)}(t)}{M_\nu}\nabla_\nu C_l^{(I)}(t)\nonumber\\
-\sum_{\nu=1}^{N_n}\frac{\mathbf P_\nu(t)+i\boldsymbol{\mathcal P}_\nu^{(I)}(t)}{M_\nu}\sum_k C_k^{(I)}(t)\mathbf d_{\nu,lk}^{(I)}.
\end{align}
This preserves the norm of the electronic wave function along the time 
evolution. On the right-hand-side of Eq.~(\ref{eqn: apx electronic eqn}), the 
terms not containing $\boldsymbol{\mathcal P}_\nu^{(I)}(t)$ are exactly the 
same as in other algorithms, i.e. Ehrenfest and TSH. The additional terms 
follow from the exact factorization and are all proportional 
to $\boldsymbol{\mathcal P}_\nu^{(I)}(t)$, the so-called quantum momentum, 
that will be discussed below. Moreover, since the coefficients in the 
expansion~(\ref{eqn: BO expansion}) depend on nuclear positions, spatial 
derivatives of such coefficients need to be taken into account.

\subsection{Spatial dependence of the coefficients of the Born-Huang expansion}
The coefficients $C_l^{(I)}(t)$ of the Born-Huang expansion of the electronic wave function are written in terms of the their modulus $\left|C_l^{(I)}(t)\right|$ and phase $\gamma_l^{(I)}(t)$,
\begin{align}
 \nabla_{\nu}C_l^{(I)}(t) &= \left[\frac{\nabla_{\nu}\left|C_l^{(I)}(t)\right|}{\left|C_l^{(I)}(t)\right|}+ \frac{i}{\hbar}\nabla_\nu\gamma_l^{(I)}(t)\right] C_l^{(I)}(t).
 \label{eqn: derivative of C_l with modulus and phase}
\end{align}
The \textit{third approximation} introduced to derive the CT-MQC algorithm consists in (i) neglecting the first term on the right-hand-side of the above equation and (ii) neglecting all terms depending on the NACVs in the expression of the remaining term. Based on the analysis reported in previous work~\cite{Gross_PRL2013, Gross_MP2013, Gross_JCP2015}, the contribution of the first term is indeed negligible if compared to the spatial derivative of the phase $\gamma_l^{(I)}(t)$. We will present our argument below and we will give the expression of the remaining term. A semiclassical analysis is provided in Section~SI.2 of the Supporting Information.

Based on the solution of the full TDSE~\cite{Gross_PRL2013, Gross_MP2013, 
Gross_JCP2015} for one-dimensional systems, we have observed that at a given 
time the quantities $\left|C_l(\dulR,t)\right|$ are either constant functions 
of $\dulR$ or present a sigmoid shape~\cite{Gross_JCP2015}. In particular, 
this second feature appears when the nuclear wave packet splits after having 
crossed a region of strong coupling. In regions where the $\left|C_l(\dulR,t)\right|$'s are constant, their 
derivatives are zero, thus our approximation holds perfectly; in the sigmoid 
case, $\left|C_l(\dulR,t)\right|$ is constant, i.e. either 0 or 1, far from 
the step of the sigmoid function, and linear around the center of the step. 
Furthermore, the center of the step is the position where the nuclear density 
splits~\cite{Gross_PRL2013}, and only in this region the gradient of
$\left|C_l(\dulR,t)\right|$ can be significantly different from zero. We have 
analytically demonstrated that these properties are generally valid~\cite{Gross_JCP2015} in the absence of external time-dependent 
fields. It is important to keep in mind that the nuclear density is 
reconstructed from the distribution of classical trajectories, thus it seems 
reasonable to assume that when the nuclear density splits into two or more 
branches, the probability of finding trajectories in the tail regions is very 
small. It follows that $\nabla_\nu |C_l^{(I)}|/|C_l^{(I)}|=0$ is an 
approximation only for few trajectories, located in the tail regions, while it 
is true for all other trajectories. Therefore, we expect that such an 
approximation will not affect drastically the final (averaged over all 
trajectories) results.

We now come back to the expression of $\nabla_\nu\gamma_l^{(I)}(t)$. First of all, let us show the time derivative of $\gamma_l^{(I)}(t)$, namely $\dot\gamma_l^{(I)}(t)=-\epsilon_{BO}^{(l),(I)}+\mbox{non-adiabatic terms}$. The derivation of the additional ``non-adiabatic terms'', containing the NACVs is tedious and only requires the knowledge of Eq.~(\ref{eqn: apx electronic eqn}). If we assume that the NACVs are localized in space, that is an approximation which is valid in many physically relevant situations~\cite{tapaviczaPRL2007, robbJPCA2003, hynes_FD2004, marxPRL2002}, we can simply write $\nabla_\nu\dot\gamma_l^{(I)}(t)=-\nabla_\nu\epsilon_{BO}^{(l)}$ or, equivalently,
\begin{align}\label{eqn: nabla gamma}
 \nabla_\nu\gamma_l^{(I)}(t)=-\int^t dt'\nabla_\nu\epsilon_{BO}^{(l),(I)}=\mathbf f_{l,\nu}^{(I)}(t).
\end{align}
Therefore, the approximate form of the spatial derivative of the coefficient 
of the Born-Huang expansion becomes 
$\nabla_{\nu}C_l^{(I)}(t) =(i/\hbar)\mathbf f_{l,\nu}^{(I)}(t)C_l^{(I)}(t)$. 
The quantity $\mathbf f_{l,\nu}^{(I)}(t)$, the time-integrated adiabatic 
force, can be determined by only computing electronic adiabatic properties. It 
follows that the electronic evolution equation~(\ref{eqn: apx electronic eqn}) 
is no longer a partial differential equation, but rather an ordinary differential equation.

Related to the discussion just presented, we can now give the explicit expression of the TDVP in terms of the adiabatic electronic properties, namely
\begin{align}
 \bA_\nu^{(I)}(t) = \sum_{l=1}^{N_{st}}\rho_{ll}^{(I)}(t)\mathbf f_{l,\nu}^{(I)}(t)
 +\hbar \operatorname{Im}\sum_{l,k=1}^{N_{st}}\rho_{lk}^{(I)}(t)\mathbf d_{\nu,lk}^{(I)},\label{eqn: vector potential with f}
\end{align}
with $\rho_{lk}^{(I)}(t)=C_l^{(I)*}(t)C_k^{(I)}(t)$ the elements of the electronic density matrix.

\subsection{Quantum momentum}
The crucial difference between standard Ehrenfest-type approaches and our new 
quantum-classical algorithm for non-adiabatic dynamics is the appearance of 
the quantum momentum in the equations of motion. In the framework of the exact 
factorization, the electronic equation~(\ref{eqn: exact electronic eqn}) contains the exact coupling to the nuclei, which is expressed in terms of $-i\hbar\nabla_\nu\chi/\chi$. As in Eq.~(\ref{eqn: quantum momentum 1}), this dependence on the nuclear wave function is written as
\begin{align}
\frac{-i\hbar\nabla_\nu\chi^{(I)}(t)}{\chi^{(I)}(t)}+\mathbf A_\nu^{(I)}(t) =\mathbf P_\nu^{(I)}(t) + i\boldsymbol{\mathcal P}_\nu^{(I)}(t).
\end{align}
In writing the first term on the right-hand-side, we are using the first approximation introduced above, where the gradient of the phase of the nuclear wave function (see Eq.~(\ref{eqn: quantum momentum 1}) for comparison) is truncated at the zero-th order term in the $\hbar$-expansion. The second term is the quantum momentum, related to the spatial variation of the nuclear density (notice that the relation $\nabla_\nu|\chi|/|\chi| = \nabla_\nu|\chi|^2/(2|\chi|^2)$ holds). This term, which has the dimension of a momentum and is purely imaginary, is a known quantity in the context of Bohmian mechanics~\cite{Garashchuk_CPL2003}.
When presenting the numerical results, it will become clear that the quantum momentum is the source of decoherence effects on electronic dynamics.

In the nuclear equation~(\ref{eqn: HJE}) we have discarded the quantum potential, while in the electronic equation~(\ref{eqn: apx electronic eqn}) we have taken the quantum momentum into account. On the one hand, the quantum potential appears $\mathcal O(\hbar^2)$, while the correction to the nuclear momentum is $\mathcal O(\hbar)$, thus our approximations are consistent up to within this order in Planck's constant. On the other hand, we have already shown~\cite{Gross_JCP2014} that neglecting the quantum momentum in the electronic equation from the exact factorization produces an Ehrenfest-like evolution, where the spatial splitting of the nuclear wave packet cannot be captured. The introduction of the quantum momentum is thus essential. Further studies on incorporating the quantum potential in Eq.~(\ref{eqn: HJE}) following the strategies such as the ones discussed in the context of Bohmian dynamics~\cite{tavernelliJCP2013, tavernelliPCCP2011} are indeed a route to be investigated.

The calculation of the quantum momentum requires calculation of the spatial 
derivatives of the nuclear density, which is known numerically only at the 
positions of the classical trajectories. Therefore, in order to evaluate such 
derivative at the position $\dulR^{(I)}(t)$, non-local information about the 
position of neighboring trajectories $\dulR^{(J)}(t)$ is necessary. This 
requirement is what makes the algorithm based on Eqs.~(\ref{eqn: force 
before gauge}) and~(\ref{eqn: apx electronic eqn}) a 
\textsl{coupled-trajectory} scheme. The procedure to calculate the quantum 
momentum used in the current implementation of the algorithm is described 
below for a two-level system, since all numerical tests are performed on such 
model situations. The extension to a multi-level system is presented in 
Section~SI.3 of the Supporting Information.

\subsubsection{Two-level system}
The nuclear density can be expressed as the sum of BO-projected densities. In Section~SI.1 of the Supporting Information it is proved that if the full wave function $\Psi$ is expanded on the adiabatic basis with coefficients $F_l(\dulR,t)$, then the factorization $\Psi=\Phi_\dulR\chi$ implies that $|\chi|^2=\sum_l|F_l|^2$ and $F_l=C_l\chi$. Therefore, in a two-level system the quantum momentum becomes
\begin{align}
\label{eqn: nabla chi/chi for two state}
\boldsymbol{\mathcal P}(\dulR,t)=\frac{-\hbar}{2}\frac{\nabla_\nu\left|\chi(\dulR,t)\right|^2}{\left|\chi(\dulR,t)\right|^2}=\frac{-\hbar}{2}\frac{\nabla_\nu\left|F_1(\dulR,t)\right|^2+\nabla_\nu\left|F_2(\dulR,t)\right|^2}{\left|F_1(\dulR,t)\right|^2+\left|F_2(\dulR,t)\right|^2}.
\end{align}
We now assume that each BO-projected nuclear wave packet is a single Gaussian. Notice that we make this approximation only to evaluate the quantum momentum, whereas the nuclear dynamics will still be represented in terms of classical trajectories. This is the \textit{fourth approximation} introduced to derive the CT-MQC algorithm. We write
\begin{align}
 \left|F_l\left(\dulR,t\right)\right|^2 = \rho_{ll}(t)
 \frac{1}{\mathcal N_l}\prod_{\nu=1}^{N_n}\exp{\left[-\frac{\left[\mathbf R_\nu-\mathbf R_\nu^{(l)}(t)\right]^2}{\sigma_l^2(t)}\right]}=\rho_{ll}(t) G_{\sigma_l}\left(\dulR-\dulR^{(l)}(t)\right)
\end{align}
where $\sigma_l(t)$ and $\mathbf R_\nu^{(l)}(t)$ are the time-dependent 
variance and mean value, respectively, of the normalized Gaussian 
$G_{\sigma_l}$. $\rho_{ll}(t)$ accounts for the normalization (the 
integral over $\dulR$ of the function $|F_l\left(\dulR,t\right)|^2$ is the 
population of the corresponding BO state; see for this Section~SI.3 of the 
Supporting Information) and $\mathcal N_l$ is the normalization constant. 
Computing the gradient of the Gaussians, the quantum momentum becomes
\begin{align}
\boldsymbol{\mathcal P}(\dulR,t) = \hbar\sum_{l=1,2} \frac{\left[\mathbf R_\nu-\mathbf R_\nu^{(l)}(t)\right]}{\sigma_l^2(t)}\left|C_l(\dulR,t)\right|^2\label{eqn: nabla chi / chi for two state}
\end{align}
and this expression can be calculated analytically if the mean position and variance are determined from the distribution of the classical trajectories, according to~\cite{Gross_JCP2015}
\begin{align}
 \mathbf R_\nu^{(l)}(t)&=\frac{1}{\rho_{ll}(t)}\int d\dulR\,\mathbf R_\nu\left|F_l\left(\dulR,t\right)\right|^2\simeq \sum_{I=1}^{N_{traj}}\mathbf R_\nu^{(I)}(t)\frac{\rho_{ll}^{(I)}(t)}{\sum_{J=1}^{N_{traj}}\rho_{ll}^{(J)}(t)}\label{eqn: mean position of the gaussian}\\
 \sigma_l^2(t)&=\frac{2}{\rho_{ll}(t)}\int d\dulR\,\big[\mathbf R_\nu-\mathbf R_\nu^{(l)}(t)\big]^2\left|F_l\left(\dulR,t\right)\right|^2\simeq\sum_{I=1}^{N_{traj}}\left[\mathbf R_\nu^{(I)}(t)-\mathbf R_\nu^{(l)}(t)\right]^2 \frac{\rho_{ll}^{(I)}(t)}{\sum_{J=1}^{N_{traj}}\rho_{ll}^{(J)}(t)}.\label{eqn: variance of the gaussian}
\end{align}
Here, we have used the following approximations
\begin{align}
 \left|F_l\left(\dulR,t\right)\right|^2&\simeq\frac{1}{N_{traj}}\sum_{I=1}^{N_{traj}}\rho_{ll}^{(I)}(t)\delta\left(\dulR-\dulR^{(I)}(t)\right)\label{eqn: approx BO wp}\\
 \rho_{ll}(t)&\simeq\frac{1}{N_{traj}}\sum_{I=1}^{N_{traj}}\rho_{ll}^{(I)}(t),\label{eqn: approx BO population}
\end{align}
with $N_{traj}$ the total number of trajectories. The first equation associates a weight, $\rho_{ll}^{(I)}(t)$, to each trajectory, in order to reconstruct the BO-projected densities from the histogram, i.e. $N_{traj}^{-1}\sum_{I}\delta(\dulR-\dulR^{(I)}(t))$, which approximates the full nuclear density. The second equation is instead used to determine the population of the adiabatic state $l$ as an average of the coefficients associated with each trajectory. The pre-factor $N_{traj}^{-1}$ stands for the weight of each trajectory, chosen to be constant. In Eqs.~(\ref{eqn: approx BO wp}) and~(\ref{eqn: approx BO population}), the integrals over the positions $\dulR$ have been replaced by the sum over trajectories since the nuclear density, for each trajectory, is a $\delta$-function centered at the position of the trajectory itself.  

An additional simplification can be introduced at this point, based on the following observations. If the quantum momentum $\boldsymbol{\mathcal P}_\nu(\dulR,t)$ is neglected in the electronic equation~(\ref{eqn: apx electronic eqn}), then the quantum-classical equations derived from the exact factorization yield Ehrenfest dynamics: the nuclear wave packet propagates coherently and its spatial splitting cannot be reproduced~\cite{Gross_JCP2014}. In this case the spatial distribution~\cite{Gross_JCP2015} of the coefficients $|C_l^{(I)}(t)|$ is not correctly reproduced, as they are more or less constant. The lack of spatial distribution is related to the lack of decoherence: the indicator used in the following analysis to quantify decoherence  (see Eq.~(\ref{eqn: indicator of decoherence})) depends in fact on the \textsl{shape} of $|C_l^{(I)}(t)|$.
Also, we have seen in previous studies~\cite{Gross_JCP2015} that steps developing in the exact TDPES, bridging regions of space where it has adiabatic shapes, is instead able to induce the splitting of the nuclear density. Outside the step region, the force is then simply adiabatic (i.e. the gradient of an adiabatic surface). The region 
of the step is where the additional force, beyond the adiabatic force, acts. 
Furthermore, the steps in the TDPES appear in the region where 
$\left|C_l(\dulR,t)\right|$, the moduli of the coefficients in the Born-Huang 
expansion~(\ref{eqn: BO expansion}), are neither 0 nor 1 (remember that 
$\left|C_l(\dulR,t)\right|$ is either constant in space or has a sigmoid 
shape between the values 0 and 1~\cite{Gross_PRL2013, Gross_MP2013, 
Gross_JCP2015}). Therefore, it seems natural to assume that an additional effect, responsible for the splitting of the nuclear density, should be localized in space in the region of the steps in the TDPES, meaning in the region 
between the two Gaussian-shaped BO densities. There, 
$\boldsymbol{\mathcal P}_\nu(\dulR,t)$ can be represented, approximately,
as a linear function, which can be determined analytically. Notice that an analogous treatment of the quantum momentum, which follows from the hypothesis that the nuclear density can be approximated as a sum of Gaussians, has already been discussed in the context of Bohmian mechanics~\cite{Garashchuk_CPL2003}. Due to this approximation, however, the coupled electronic and nuclear equations do not fulfil some fundamental properties, i.e. population exchange between electronic states might be observed even when the NACVs are zero. 

The linear function used to approximate the quantum momentum is determined by 
associating two parameters to it: a $y$-intercept, $\dulR^0$, by imposing in 
Eq.~(\ref{eqn: apx electronic eqn}) that no population variation shall be 
observed if the NACVs are zero, and a slope, $\alpha$, determined 
analytically from Eq.~(\ref{eqn: nabla chi / chi for two state}) evaluated at 
$\dulR^0$. The first parameter is determined by setting the terms 
containing the NACVs in Eq.~(\ref{eqn: apx electronic eqn}) to zero and then 
imposing that the remaining part, 
$\dot{\rho}_{ll}^{(I)} = \sum_\nu[2\boldsymbol{\mathcal P}_\nu^{(I)}/(\hbar M_\nu)]\cdot[\mathbf A_\nu^{(I)}-\mathbf f_{l,\nu}^{(I)}]\rho^{(I)}_{ll}$
, has to be zero when summed up over the trajectories. When imposing this 
condition, the expression used for the quantum momentum is simply 
$\boldsymbol{\mathcal P}_\nu^{(I)}=\alpha(\mathbf R_\nu^{(I)}-\mathbf R^0_\nu)$. 
As indicated above, once the $y$-intercept is known, it is inserted into 
Eq.~(\ref{eqn: nabla chi / chi for two state}) and the slope is obtained 
analytically, yielding the quantum momentum as
\begin{align}\label{eqn: QM}
\boldsymbol{\mathcal P}_\nu^{(I)}(t) = \hbar\left( \sum_{l=1,2}\frac{\left|C_l\left(\dulR,t\right)\right|^2\bigg|_{\mathbf R_\nu^0(t)}}{\sigma^2_l(t)}\right)\left[\mathbf R_\nu^{(I)}(t)-\sum_{J}\mathbf R_\nu^{(J)}(t)\frac{\rho^{(J)}_{11}(t)\rho^{(J)}_{22}(t)\left(\mathbf f_{1,\nu}^{(J)}(t)-\mathbf f_{1,\nu}^{(J)}(t)\right)}{\sum_{J}\rho^{(J)}_{11}(t)\rho^{(J)}_{22}(t)\left(\mathbf f_{1,\nu}^{(J)}(t)-\mathbf f_{1,\nu}^{(J)}(t)\right)}\right]
\end{align}
This expression is only used in the region between the centers of the BO-projected densities, $\mathbf R_\nu^{(l)}(t)$. Outside this region, the quantum momentum is set zero.

\section{Numerical Implementation}\label{sec: flowchart}
In this section, we present the numerical implementation of the CT-MQC algorithm.
First, we summarize the final expressions of the classical force, used to generate the trajectories, and the evolution equation for the coefficients of the Born-Huang expansion of the electronic wave function. The CT-MQC equations are cast below in such a way that the first line in both expressions is exactly the same as in Ehrenfest-like approaches, while the corrections, in the second line of each expression, are (i) proportional to the quantum momentum, (ii) do not contain the NACVs, i.e. the ``competing'' effects of population exchange induced by the NACVs and of decoherence induced by the quantum momentum have been separated (this is the \textit{fifth approximation} introduced in the equations~(\ref{eqn: exact electronic eqn}) and~(\ref{eqn: exact nuclear eqn}) to derive the CT-MQC algorithm). The classical force
\begin{align}\label{eqn: final equation for the classical force}
\dot{\mathbf P}_\nu^{(I)}(t)&=-\sum_{k}\rho_{kk}^{(I)}(t)\nabla_\nu\epsilon_{BO}^{(k),(I)}-\sum_{k,l}\rho_{lk}^{(I)}(t) \left(\epsilon_{BO}^{(k),(I)}-\epsilon_{BO}^{(l),(I)}\right)
\mathbf d_{\nu,lk}^{(I)} \\
&-\sum_{l}\rho_{ll}^{(I)}(t)\left(\sum_{\nu'=1}^{N_n}\frac{2}{\hbar M_{\nu'}}\boldsymbol{\mathcal P}_{\nu'}^{(I)}(t)\cdot\mathbf f_{l,\nu'}^{(I)}(t)\right)\left[\sum_k \left|C_k^{(I)}(t)\right|^2\mathbf f_{k,\nu}^{(I)}(t)-\mathbf f_{l,\nu}^{(I)}(t)\right],\nonumber  
\end{align}
is used in Hamilton's equations, providing positions and momenta at each time, thus yielding trajectories in phase-space. The velocity-Verlet algorithm is used to integrate Hamilton's equations. The ordinary differential equation for the evolution of the coefficients in the electronic wave function expansion is
\begin{align}\label{eqn: final electronic eqn}
\dot C_l^{(I)}(t) &= \frac{-i}{\hbar}\epsilon_{BO}^{(l)(I)}C_l^{(I)}(t)-\sum_k C_k^{(I)}(t)\sum_{\nu=1}^{N_n}\frac{\mathbf P_\nu(t)}{M_\nu}\cdot\mathbf d_{\nu,lk}^{(I)}\\
&-\sum_{\nu=1}^{N_n}\frac{\boldsymbol{\mathcal P}_\nu^{(I)}(t)}{\hbar M_\nu}\cdot\left[\sum_k \left|C_k^{(I)}(t)\right|^2\mathbf f_{k,\nu}^{(I)}(t)-\mathbf f_{l,\nu}^{(I)}(t) \right]C_l^{(I)}(t),\nonumber
\end{align}
which is integrated using a fourth-order Runge-Kutta algorithm. 

\begin{figure}[h!]
 \begin{center}
  \includegraphics[width=0.48\textwidth]{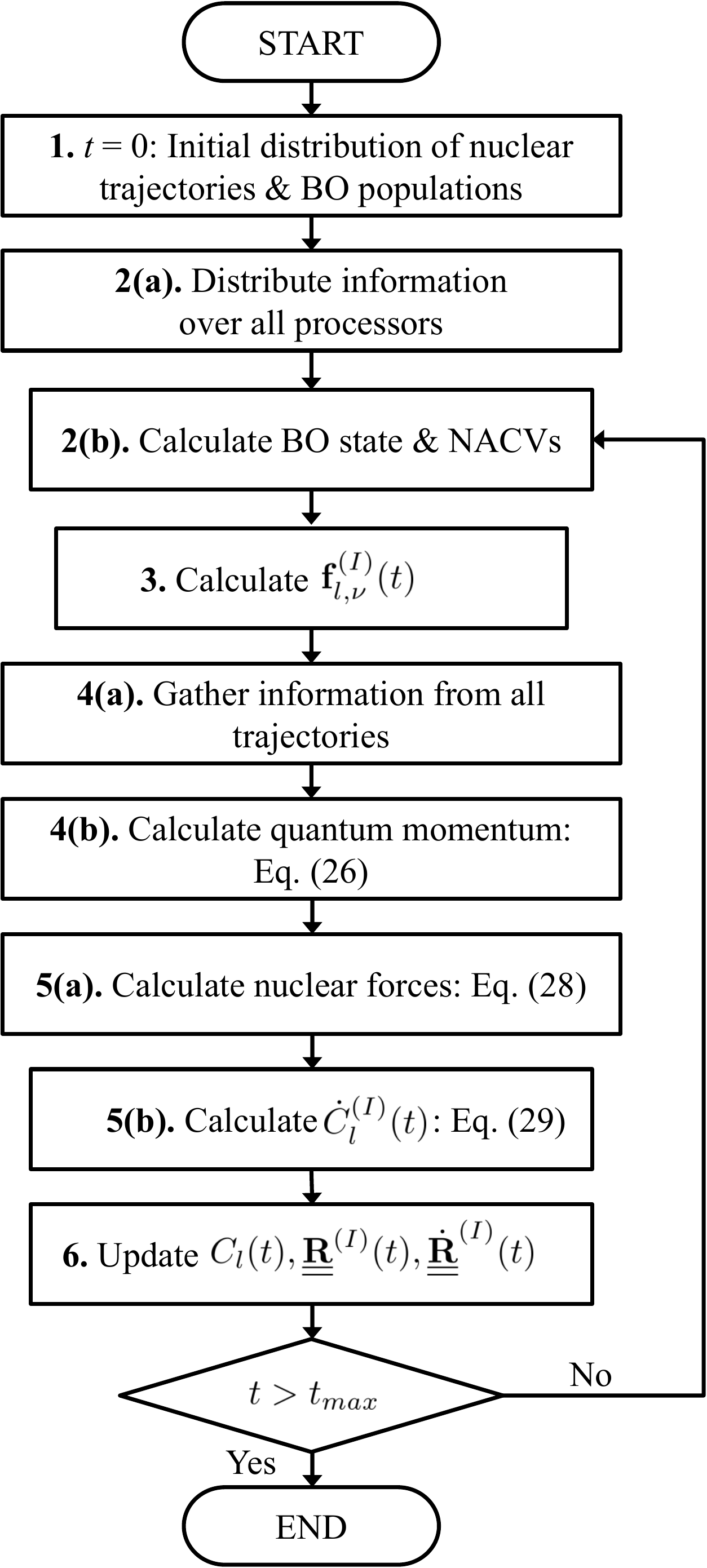}
 \end{center}
 \caption{Flowchart describing the numerical implementation of the CT-MQC.}
 \label{fig: flowchart}
\end{figure}

As stated above, the CT-MQC equations of motion are similar to Ehrenfest equations, apart from the decoherence terms, i.e. those proportional to $\boldsymbol{\mathcal P}_\nu^{(I)}(t)$, in both the nuclear and electronic equations. There, the quantum momentum is determined at each time $t$ using a multiple-trajectory scheme. Therefore, the time integration has to be performed for all trajectories simultaneously, leading to a procedure slightly different from the ``traditional'' independent-trajectory approach of the TSH method.   
In Fig.~\ref{fig: flowchart}, we give a schematic description of the steps to be performed to implement the CT-MQC algorithm, namely:
\begin{enumerate}
\item we select a set of nuclear positions and momenta, and the initial running BO state. The initial phase-space distribution can be obtained either by constructing the Wigner distribution corresponding to an initial quantum nuclear wave packet (as done in the results reported below), or by sampling the Boltzmann distribution at a given temperature, for instance via a molecular dynamics run in the canonical ensemble;
\item (a) the information about the nuclear positions is distributed on multiple processors, and (b) for each trajectory the static electronic Schr\"odinger equation is solved to obtain BO potential energies $\epsilon^{(l),(I)}_{BO}$, their gradients $\nabla_\nu\epsilon^{(l),(I)}_{BO}$, and NACVs among the BO states ${\bf d}^{(I)}_{\nu,lk}$;
\item we compute ${\bf f}^{(I)}_{l,\nu}(t)$ by accumulating up to time $t$ the BO force, $-\nabla_\nu\epsilon^{(l),(I)}_{BO}$;
\item (a) we gather the information about all nuclear trajectories $\dulR^{(I)}$, BO populations $\rho^{(I)}_{ll}$, and ${\bf f}^{(I)}_{l,\nu}$ to (b) compute the quantum momentum from Eq.~(\ref{eqn: QM});
\item (a) we calculate the nuclear force and (b) the time derivative of BO coefficients according to Eqs.~(\ref{eqn: final equation for the classical force}) and (\ref{eqn: final electronic eqn}), respectively, for each nuclear trajectory;
\item we perform the time integration for both trajectories and BO coefficients, to get the initial conditions for the following time-step. We repeat the procedure starting from point 2. until the end of the simulation.
\end{enumerate}
Step 4. in the implementation is what distinguishes the CT-MQC approach from ``independent'' multiple-trajectory methods. For all trajectories, the time integration has to be performed simultaneously, since the BO populations, the positions of the trajectories and the time-integrated BO forces have to be shared among all trajectories to compute the quantum momentum. The parallelization of the algorithm is thus essential for numerical efficiency.

One of the advantages of the CT-MQC algorithm compared to the TSH algorithm is that it is not stochastic. The electronic equation yields the proper population of the BO states,\footnote{If the squared moduli of the coefficients are multiplied by the nuclear density and integrated over nuclear space, as proven in Section~SI.1 of the Supporting Information.} including decoherence effects, since it has been derived from the exact factorization equations. Therefore, even a small number of trajectories is able to provide reliable and accurate results. The (computational) bottleneck of the algorithm is that the electronic properties (BO energies, NACVs and BO forces) are necessary at each time for all adiabatic states, whereas in TSH only ``running state'' information and the scalar product between the nuclear velocity and the NACVs are necessary.

\section{Numerical tests}\label{sec: results}
Numerical results will be presented for one-dimensional model systems~\cite{tully1990, subotnikJCP2011_1} that have nowadays become standard tests for any new quantum-classical approach to deal with non-adiabatic problems. 
Results of the CT-MQC are compared not only with an exact wave packet propagation scheme, but also with other quantum-classical approaches, namely Ehrenfest~\cite{ehrenfest, tully_fardisc1998, drukker}, TSH~\cite{tully1990, tully1971, tully1998} and independent-trajectory MQC~\cite{Gross_EPL2014, Gross_JCP2014} (the zero-th order version of the algorithm proposed here). It will be clear from the results that the use of coupled trajectories, rather than independent trajectories, which all approaches but CT-MQC are based on, is indeed the key to electronic decoherence in non-adiabatic processes.

\begin{figure}[h!]
 \begin{center}
  \includegraphics[width=0.8\textwidth,angle=270]{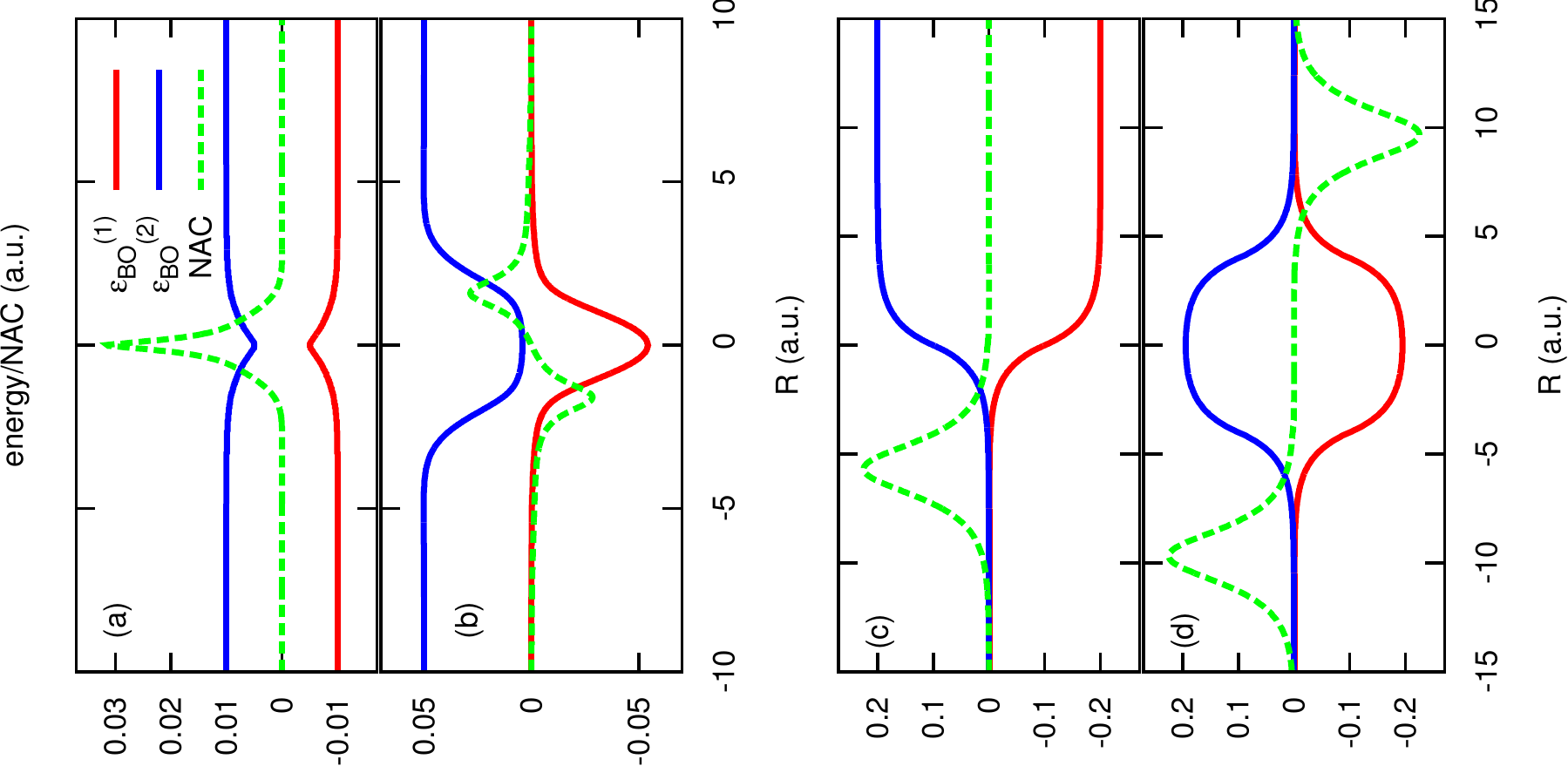}
 \end{center}
 \caption{BOPESs (first, red line, and second, blue line) and NACVs (dashed green line) for the four model systems studied in the paper. The panels correspond to the problems of (a) single avoided crossing, (b) dual avoided crossing, (c) extended coupling region with reflection and (d) double arch. In panel (a) the curve representing the NACV has been reduced 50 times, in (b) 30 times. All quantities are shown in atomic units (a.u.).}
 \label{fig: BOPES}
\end{figure}

The models discussed below are (a) single avoided crossing, (b) dual avoided 
crossing, (c) extended coupling region with reflection, (d) double arch. The 
diabatic Hamiltonians are presented in Section~SI.4 of the Supporting 
Information, while adiabatic PESs and NACVs are shown in 
Fig.~\ref{fig: BOPES}.

In presenting the numerical results, particular attention is devoted to decoherence. In order to quantify decoherence, we use as indicator
\begin{align}\label{eqn: indicator of decoherence}
\left|\rho_{12}(t)\right|^2 = \frac{1}{N_{traj}}\sum_{I=1}^{N_{traj}} \left|C_1^{(I)}(t)\right|^2\left|C_2^{(I)}(t)\right|^2,
\end{align}
which is an average over the trajectories of the (squared moduli of the) off-diagonal elements of the electronic density matrix in the adiabatic basis. The corresponding quantum mechanical quantity is
\begin{align}\label{eqn: coherence}
 \left|\rho_{12}(t)\right|^2 = \int d\dulR\,\left|C_1\left(\dulR,t\right)\right|^2\left|C_2\left(\dulR,t\right)\right|^2
\left|\chi\left(\dulR,t\right)\right|^2,
\end{align}
as proven in Section SI.4 of the Supporting Information.

\subsection{(a) - Single avoided crossing}
A Gaussian wave packet is prepared on the lower adiabatic surface and launched from the far negative region with positive initial momentum towards the avoided crossing. Two values of the mean momentum are shown in the figures below, namely $\hbar k_0=10$~a.u. and $\hbar k_0=25$~a.u.
\begin{figure}[h!]
 \begin{center}
  \includegraphics[width=0.4\textwidth]{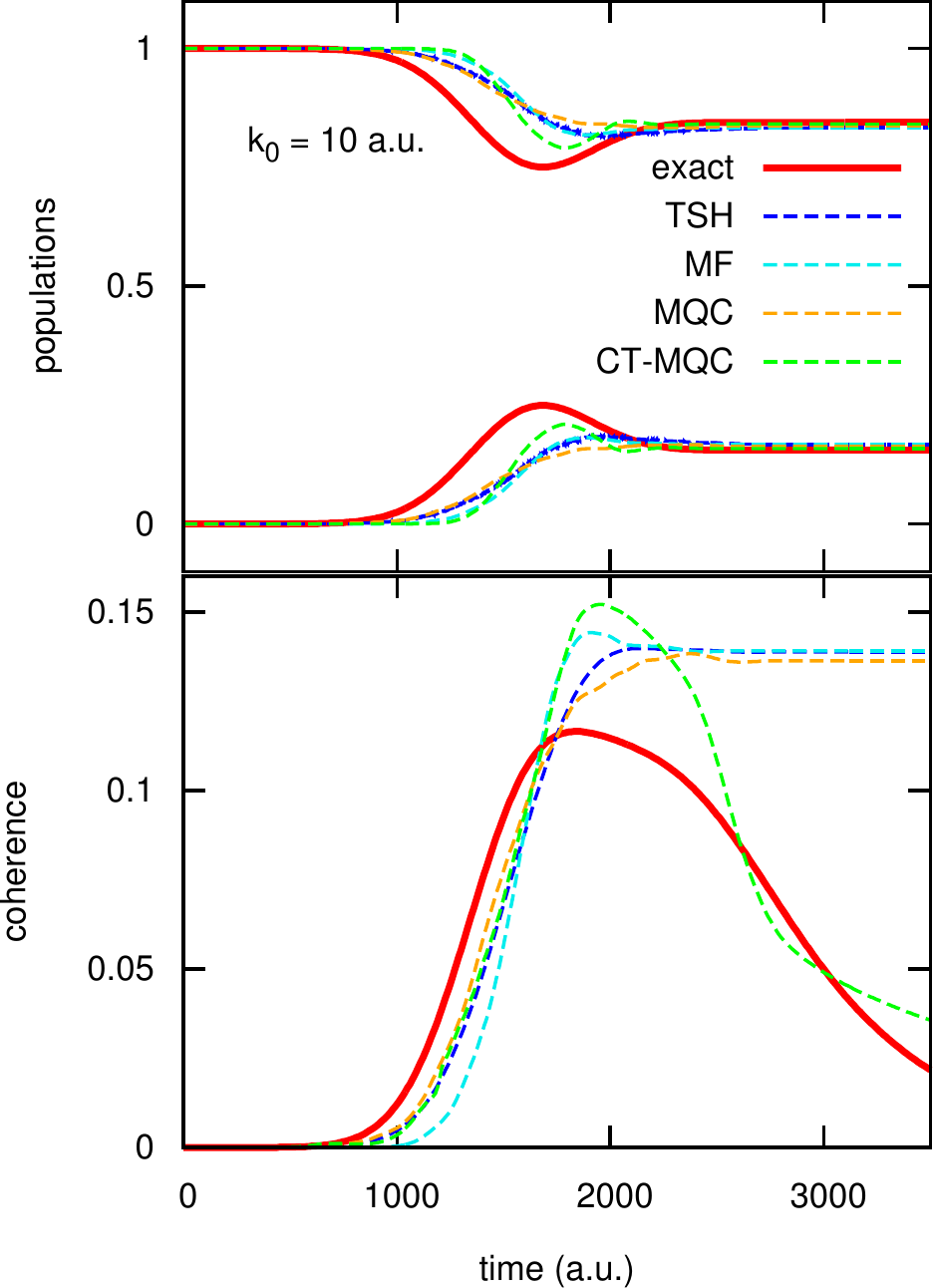}
  \includegraphics[width=0.4\textwidth]{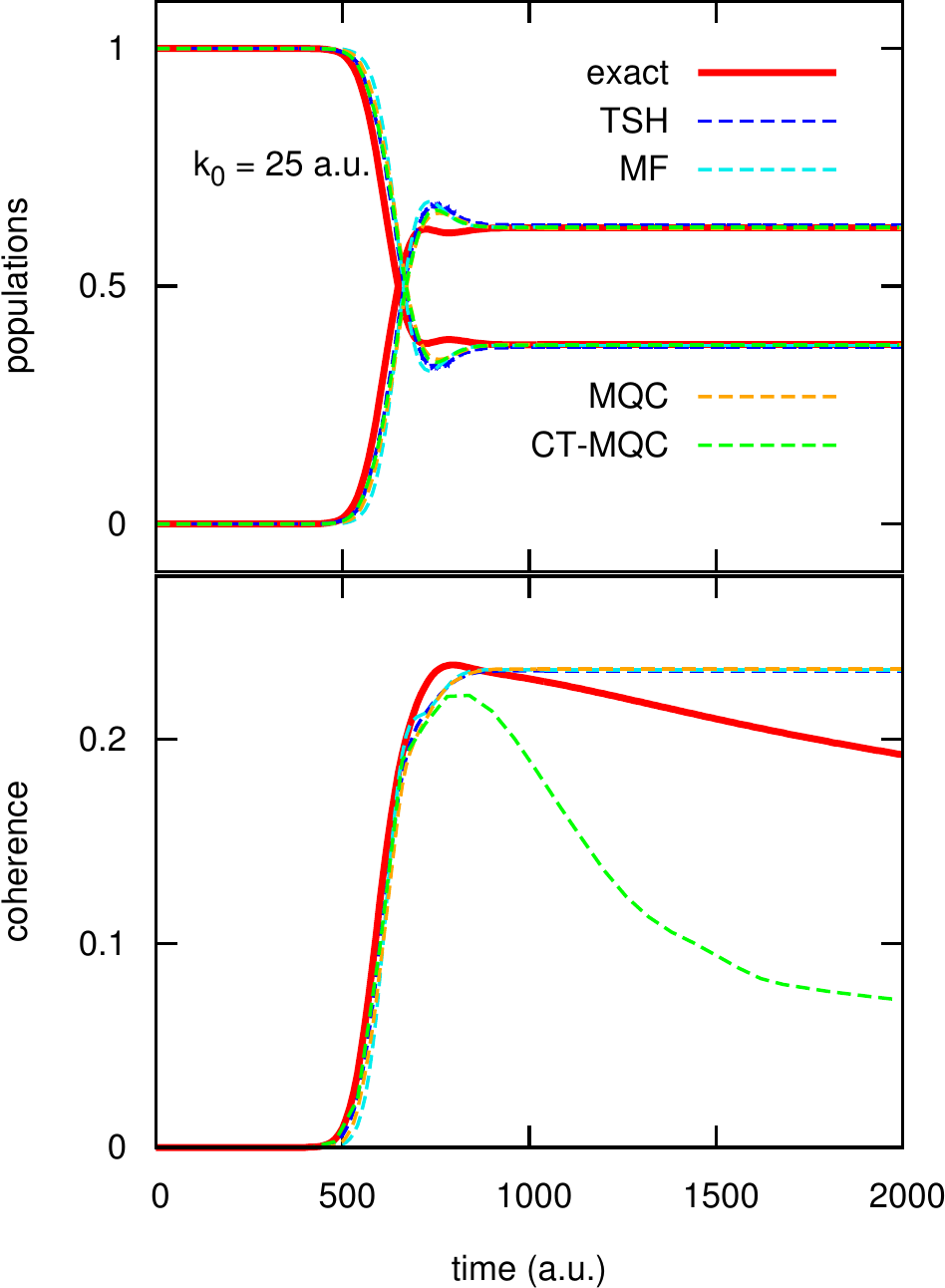}
 \end{center}
 \caption{(Left) Upper panel: populations of the BO states as functions of time, computed via exact wave packet dynamics (red) and via the approximate schemes, TSH (blue), Ehrenfest MF (cyan), MQC (orange) and CT-MQC (green), for model (a). Lower panel: indicator of decoherence as function of time. The color code is the same as in the upper panel. The results are shown for model (a) using the value $\hbar k_0=10$~a.u. for the initial momentum. (Right) Same as in the left panels but for the value $\hbar k_0=25$~a.u. of the initial momentum.}
 \label{fig: model1a}
\end{figure}
We have computed the populations of the adiabatic states as functions of time 
(upper panels in Fig.~\ref{fig: model1a}), along with the indicator for the 
decoherence (lower panels in Fig.~\ref{fig: model1a}), given in Eq.~(\ref{eqn: indicator of decoherence}). Fig.~\ref{fig: 
model1a} shows those quantities for the low initial momentum (left) and for 
the high initial momentum (right).  Numerical results are shown for exact wave 
packet propagation (red lines), TSH (blue lines), Ehrenfest dynamics (mean 
field, MF, cyan line), MQC (the independent trajectory version of the 
algorithm proposed here, orange) and CT-MQC (based on Eqs.~(\ref{eqn: final 
equation for the classical force}) and~(\ref{eqn: final electronic eqn}), 
green lines). While all methods correctly predicted the electronic population 
after the passage through the avoided crossing, only the CT-MQC scheme is able 
to qualitatively account for electronic decoherence effects. CT-MQC results 
are not in perfect agreement with the exact ones, but indeed the algorithm 
captures a trend that is completely missed by the other methods.

\begin{figure}
  \includegraphics[width=0.5\textwidth,angle=270]{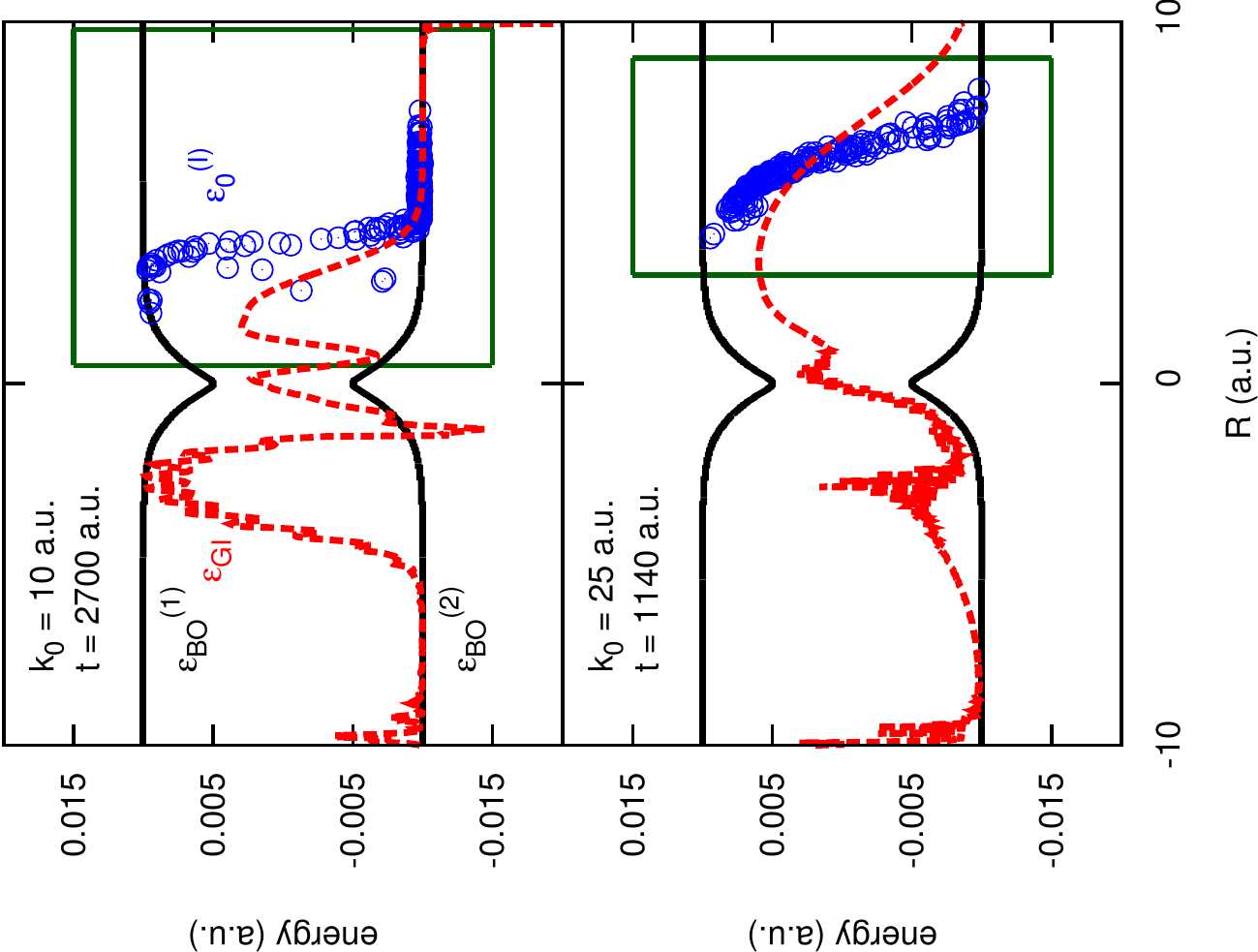}
  \includegraphics[width=0.5\textwidth,angle=270]{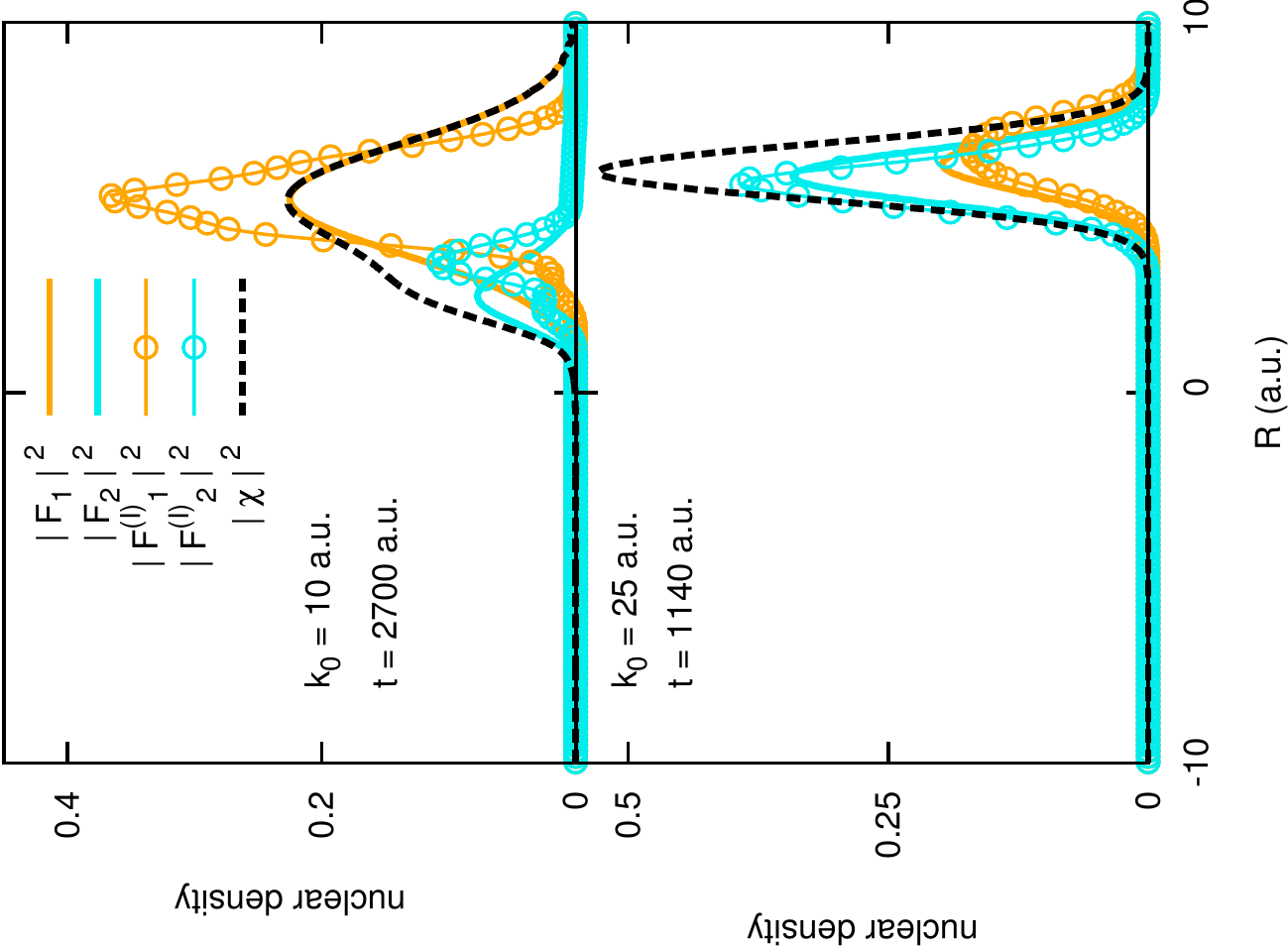}
 \caption{(Left) Snapshots (at time $t=2700$~a.u. for the low initial momentum $\hbar k_0=10$~a.u., upper panel, and at time $t=1140$~a.u. for the high initial momentum $\hbar k_0=25$~a.u., lower panel) of the gauge-invariant part of the TDPES ($\epsilon_{GI}$), shown as red lines for model (a). The black lines represent the BO surfaces, $\epsilon_{BO}^{(1)}$ and $\epsilon_{BO}^{(2)}$. The blue dots are the quantum-classical approximation to the potential. The regions highlighted in the green boxes are the regions where the calculation of the TDPES is reliable, since the nuclear density is different from zero. Outside this region the density is (numerically) zero and the results are affected by numerical noise. (Right) Snapshots at the same time-steps indicated in the left panel of the nuclear density (black lines) and of the BO-projected densities ($|F_1|^2$ orange lines and $|F_2|^2$ cyan lines) for the two initial momenta (upper panel $\hbar k_0=10$~a.u. and lower panel $\hbar k_0=25$~a.u.). The colored dots represent the quantum-classical approximation of the BO-projectes densities.}
 \label{fig: model1b}
\end{figure}

As we have derived the CT-MQC scheme from the exact factorization, we have access to the time-dependent potentials of the theory, i.e. the TDPES and the TDVP. In Fig.~\ref{fig: model1b} (left panels) we report the gauge-invariant part of the TDPES (red line), $\epsilon_{GI}$ in the figures, and we compare it with the same quantity computed with the CT-MQC procedure (blue dots), $\epsilon_0^{(I)}$ in the figures, for both initial momenta at a given time-step as indicated in the figure.
The gauge-invariant part of the TDPES is given by the first two terms in the 
definition~(\ref{eqn: tdpes}) of $\epsilon(\dulR,t)$, whose approximation in 
the quantum-classical case is simply the first term on the right-hand-side of 
Eq.~(\ref{eqn: approx TDPES}). In Fig.~\ref{fig: model1b}, the adiabatic PESs 
are shown for reference as black lines. If we observe the shape of the TDPES 
and of its approximation highlighted by the box (region where the nuclear 
density is significantly different from zero, thus it allows for the 
calculation of the TDPES), we see that the steps are reproduced very well in the 
quantum-classical picture. Since the CT-MQC correctly captures the shape of 
the potential, the nuclear density and the BO-projected densities are well 
reproduced. This is shown in the right panels in Fig.~\ref{fig: model1b}, for 
both initial momenta and at the same time-steps indicated in the left panels. 
Here, the nuclear density is indicated in black, while the BO-projected 
densities $|F_l|^2$ with $l=1,2$, are indicated as colored lines (exact) and 
dots (CT-MQC).

It is important to notice that the quantity shown in Fig.~\ref{fig: model1b}, the gauge-invariant part of the TDPES, is the only meaningful quantity that can be compared with exact results. For instance, the  TDVP, being a gauge-dependent potential, is different if different gauges are used within the exact factorization. In particular, in the exact case we have chosen to work in a gauge where the TDVP is always zero, while quantum-classical calculations are performed in the gauge defined by Eq.~(\ref{eqn: gauge}).

\subsection{(b) - Dual avoided crossing}
\begin{figure}[h!]
 \begin{center}
  \includegraphics[width=0.4\textwidth]{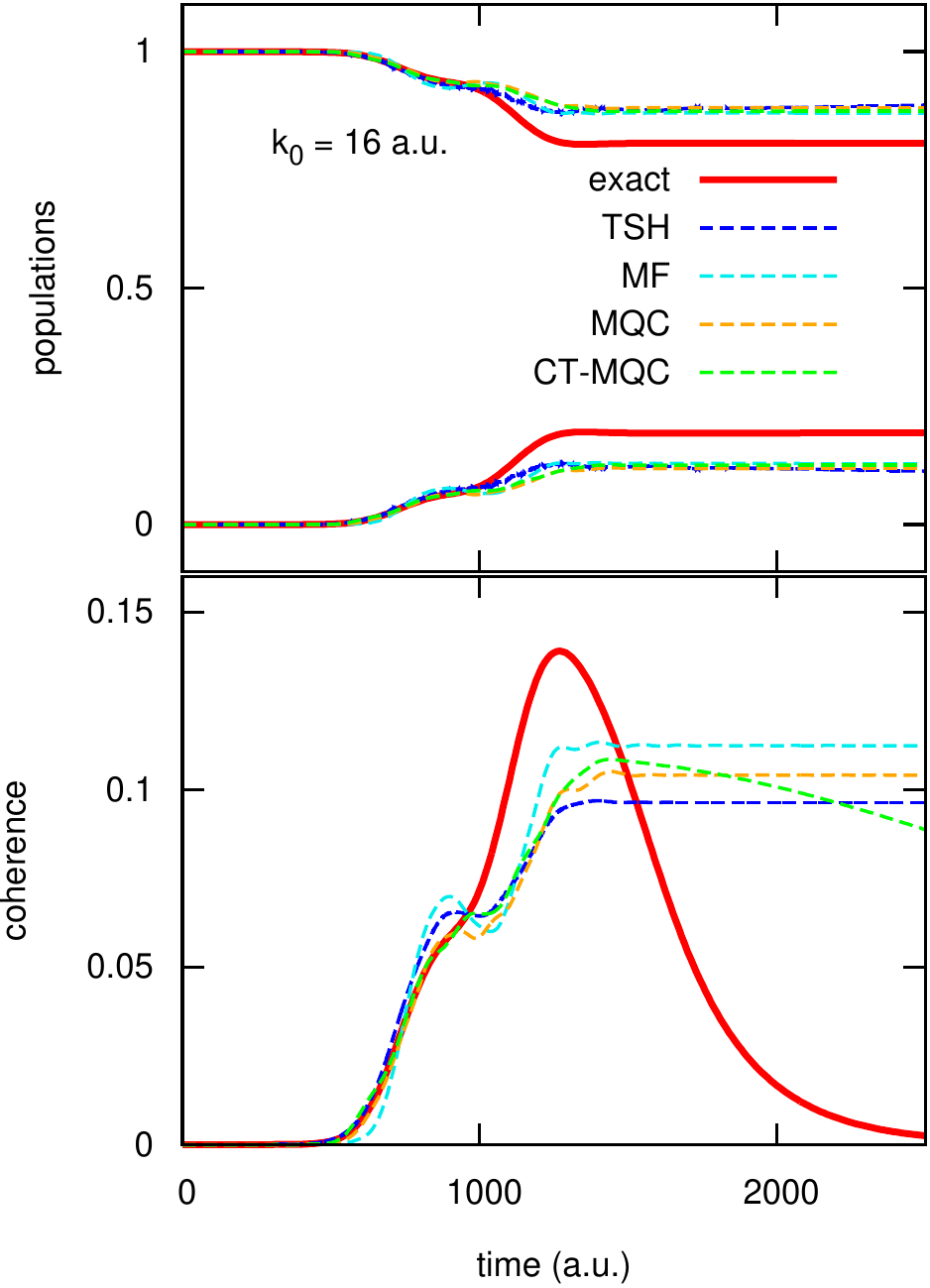}
  \includegraphics[width=0.4\textwidth]{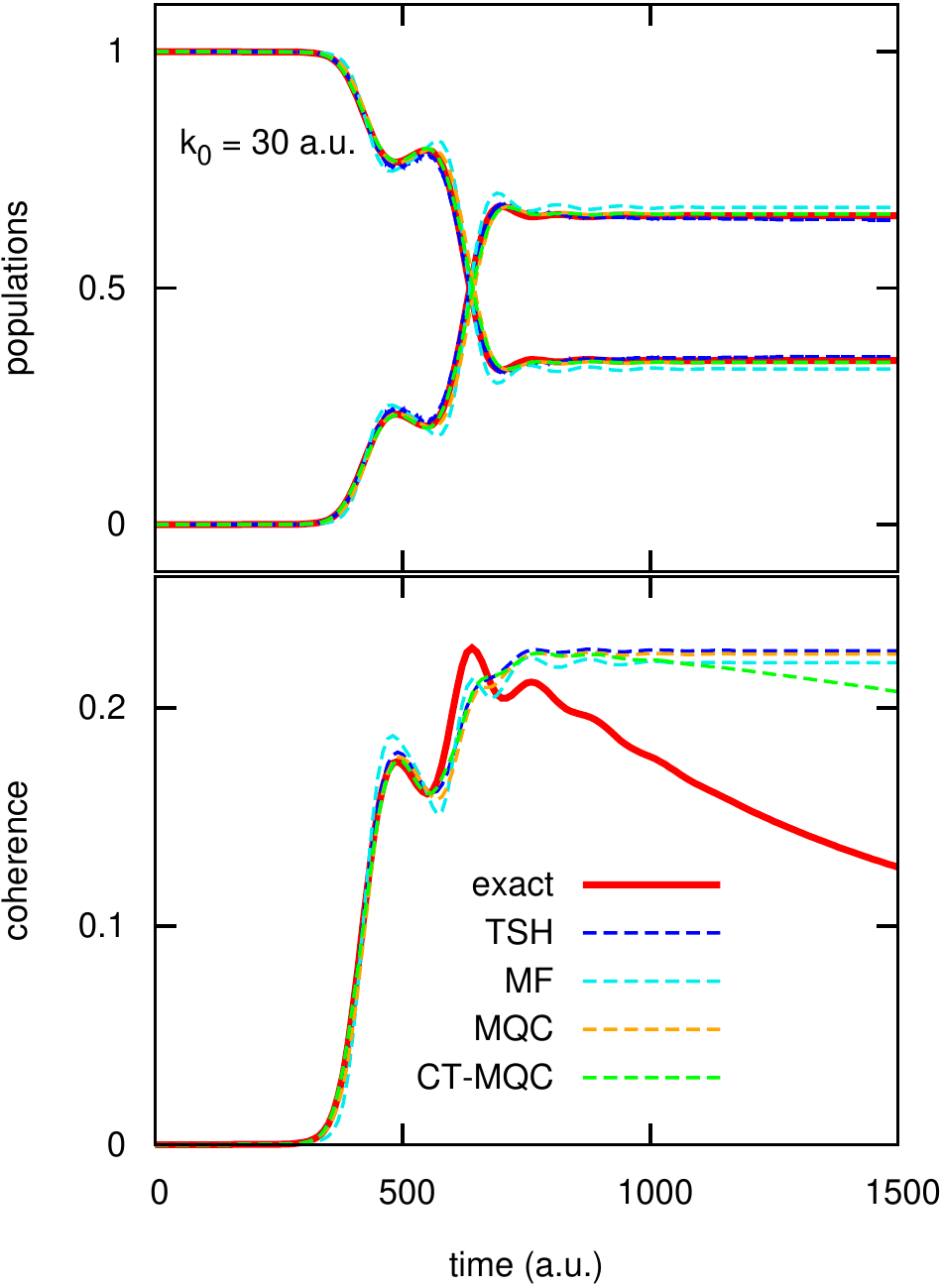}
 \end{center}
 \caption{(Left) Same as in Fig.~\ref{fig: model1a} but for the model system (b) and for the value $\hbar k_0=16$~a.u. of the initial momentum. (Right) Same as in the left panel but for the value $\hbar k_0=30$~a.u. of the initial momentum.}
 \label{fig: model2a}
\end{figure}

The same initial conditions as for model (a) have been used for the dual avoided crossing, with $\hbar k_0=16$~a.u. and $\hbar k_0=30$~a.u. chosen as initial momenta. We notice in Fig.~\ref{fig: model2a} that for low initial momentum the new algorithm is not able to correctly reproduce the final population of the adiabatic states after the two consecutive passages thorough the avoided crossings.
For both momentum values, also decoherence is not correctly captured. We 
observe however a slight deviation of the CT-MQC results from the other 
(independent-trajectory) algorithms, since the green lines in the lower panels 
of Fig.~\ref{fig: model2a} decay, but such decay is much slower than the 
expected behavior predicted by quantum mechanical results (red lines). 
In model (b) the avoided crossing regions are very close to each 
other. Therefore, the overall effect of NACVs is not localized in space, in 
constrast to what we have assumed in deriving Eq.~(\ref{eqn: nabla gamma}). Furthermore, 
in the fifth approximation introduced above we have assumed that population 
exchange due to the NACVs and decoherence due to the quantum momentum are 
separated effects. Model (b), at low initial momentum, is a situation where 
this does not happen. The nuclear wave packet encounters the first avoided 
crossing and branches on different surfaces. Then decoherence starts 
appearing, but at this point the wave packets cross the second non-adiabatic 
region. The combined effect of NACVs and quantum momentum is thus not 
completely taken into account by the CT-MQC equations. It follows that the 
CT-MQC scheme does not capture correctly the adiabatic populations. However, 
we do capture the separation of the nuclear wave packet on different BO 
surfaces, as described below.

\begin{figure}[h!]
 \begin{center}
  \includegraphics[width=0.5\textwidth,angle=270]{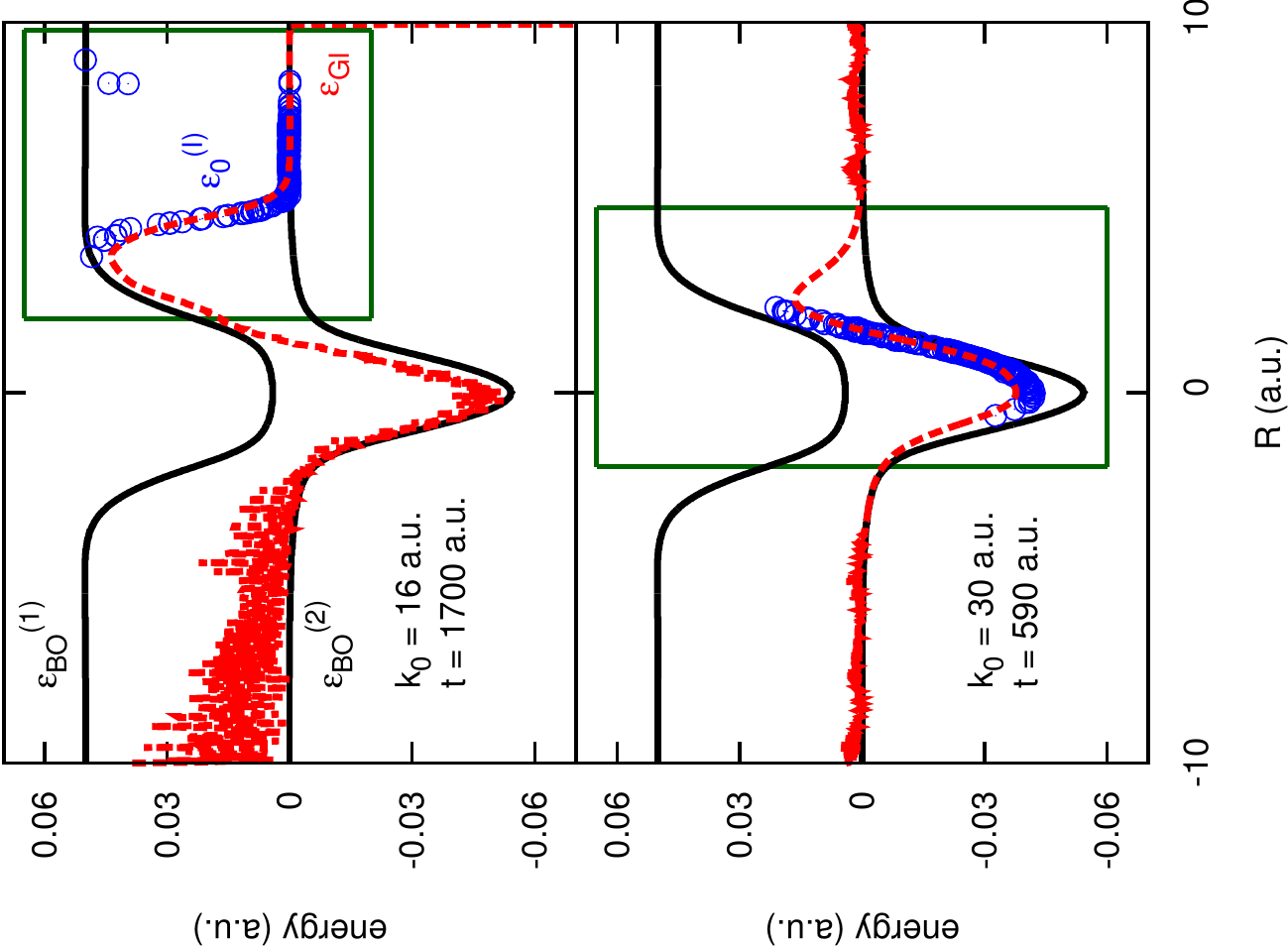}
  \includegraphics[width=0.5\textwidth,angle=270]{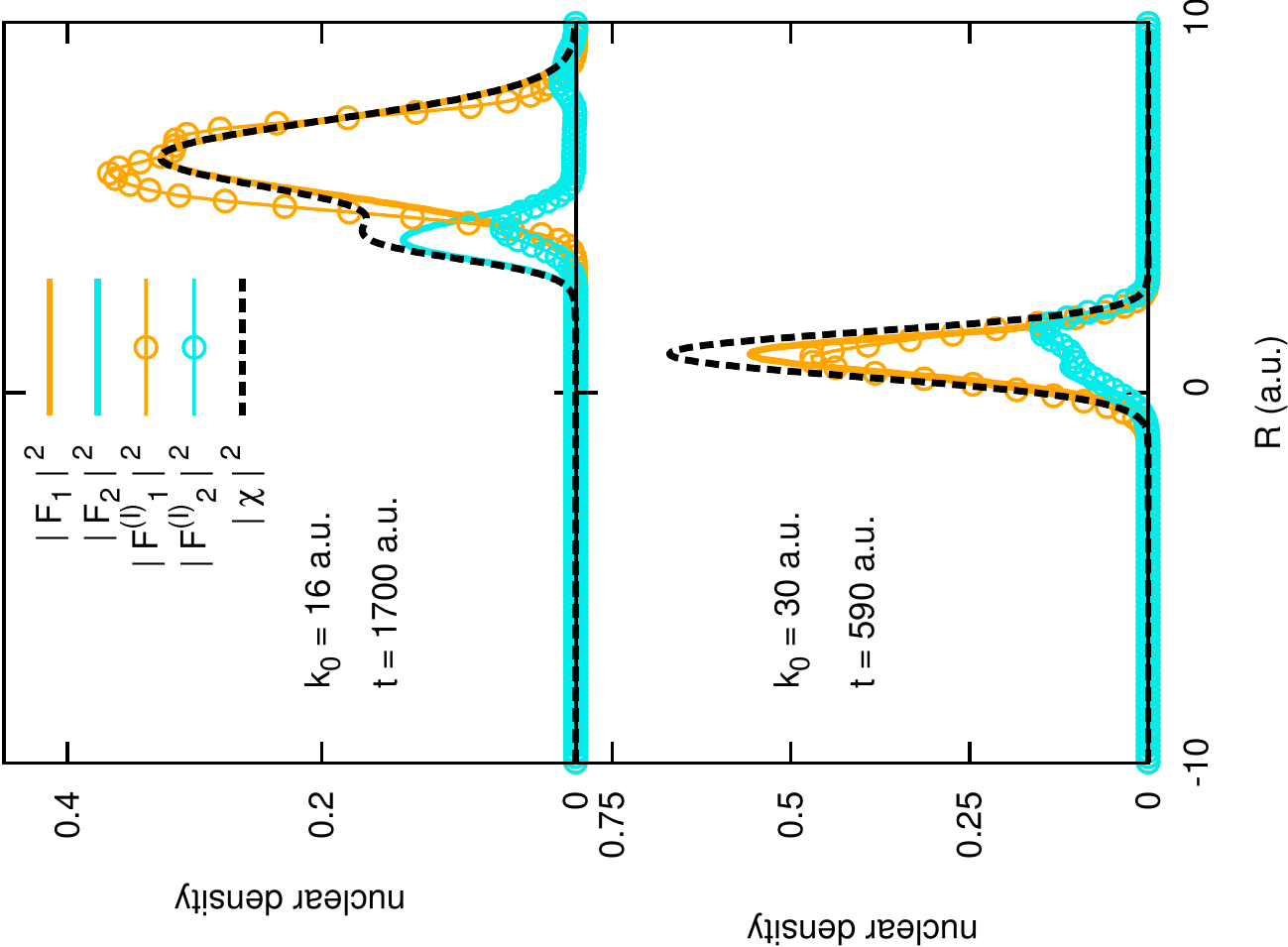}
 \end{center}
 \caption{(Left) Same as in Fig.~\ref{fig: model1b}, but the snapshots are shown at time $t=1700$~a.u. for the low initial momentum $\hbar k_0=16$~a.u. (upper panel) and at time $t=590$~a.u. for the high initial momentum $\hbar k_0=30$~a.u. (lower panel), for model (b).}
 \label{fig: model2b}
\end{figure}
Fig.~\ref{fig: model2b} shows the gauge-invariant part of the TDPES (left panels) and the nuclear densities (right panels) for both initial momenta. Despite the deviations of quantum-classical results from exact results for the electronic properties, the time-dependent potential and, consequently, the nuclear dynamics are correctly reproduced in the approximate picture.

\subsection{(c) - Extended coupling region with reflection}
This and the following model systems represent critical tests of decoherence. In both cases, the structure of the adiabatic surfaces, one with a well and one with a barrier, is responsible for yielding a high probability of reflection, especially at low initial momenta. In the case when one branch of the nuclear wave packet is reflected and the other is transmitted, the two wave packets propagate along diverging paths in nuclear space thus they lose memory of each other. This effect can be accounted for in a coupled-trajectory picture but not if independent trajectories are used, since the electronic equations are propagated fully coherently along each (independent from each other) trajectory. 

\begin{figure}[h!]
 \begin{center}
  \includegraphics[width=0.55\textwidth,angle=270]{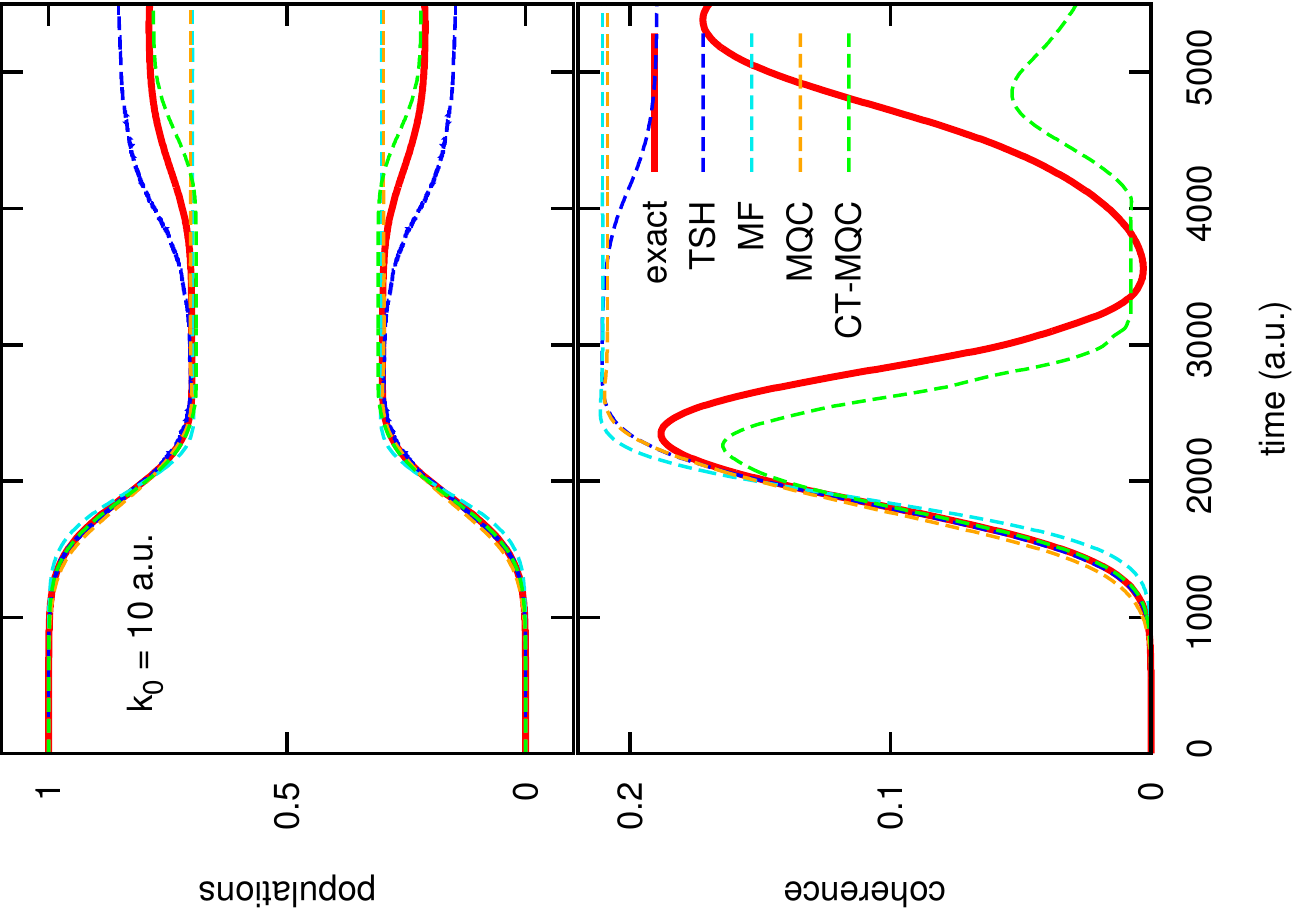}
  \includegraphics[width=0.55\textwidth,angle=270]{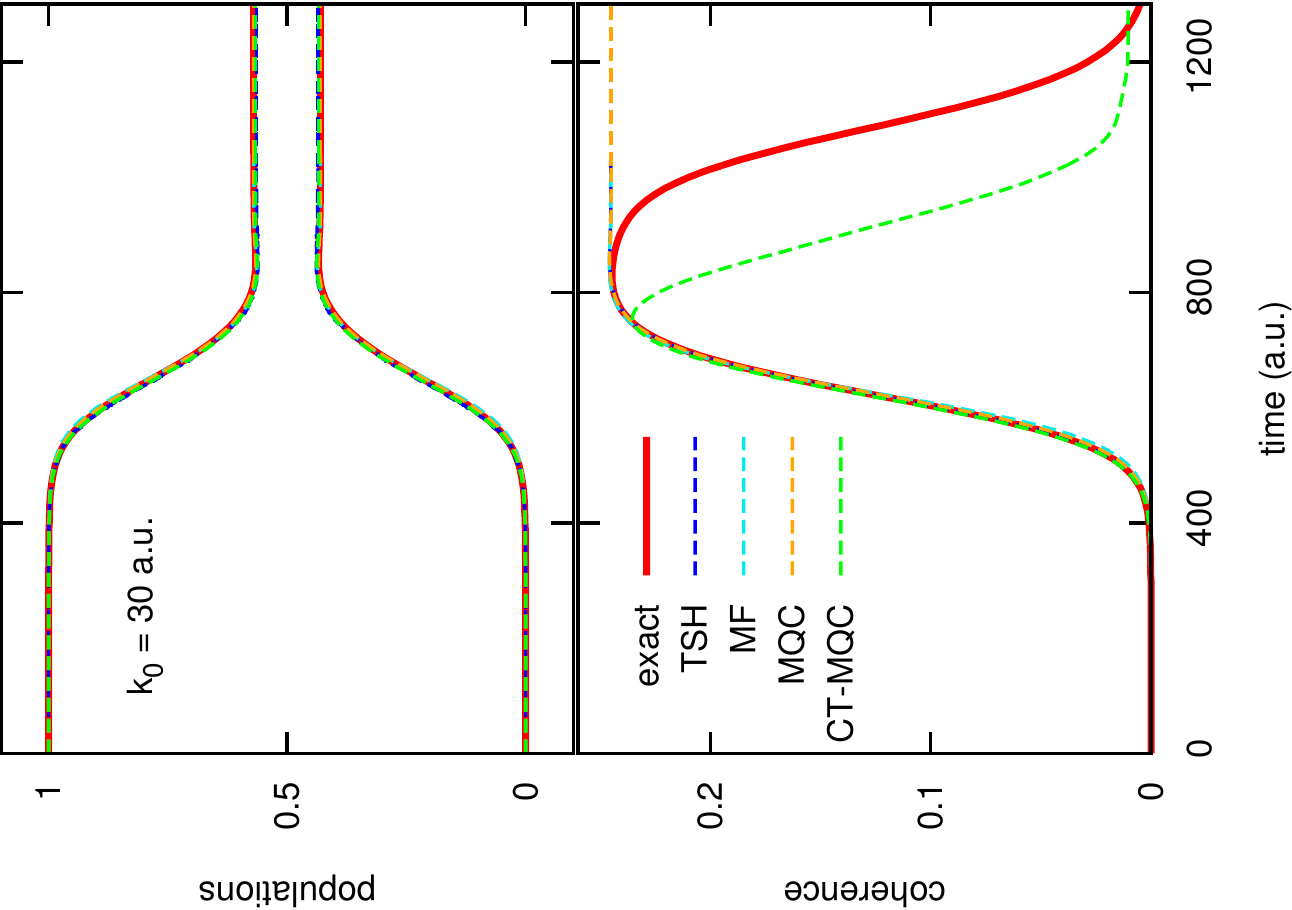}
 \end{center}
 \caption{(Left) Same as in Fig.~\ref{fig: model1a} but for the model system (c) and for the value $\hbar k_0=10$~a.u. of the initial momentum. (Right) Same as in the left panel but for the value $\hbar k_0=30$~a.u. of the initial momentum.}
 \label{fig: model3a}
\end{figure}

\begin{figure}[h!]
 \begin{center}
  \includegraphics[width=0.5\textwidth,angle=270]{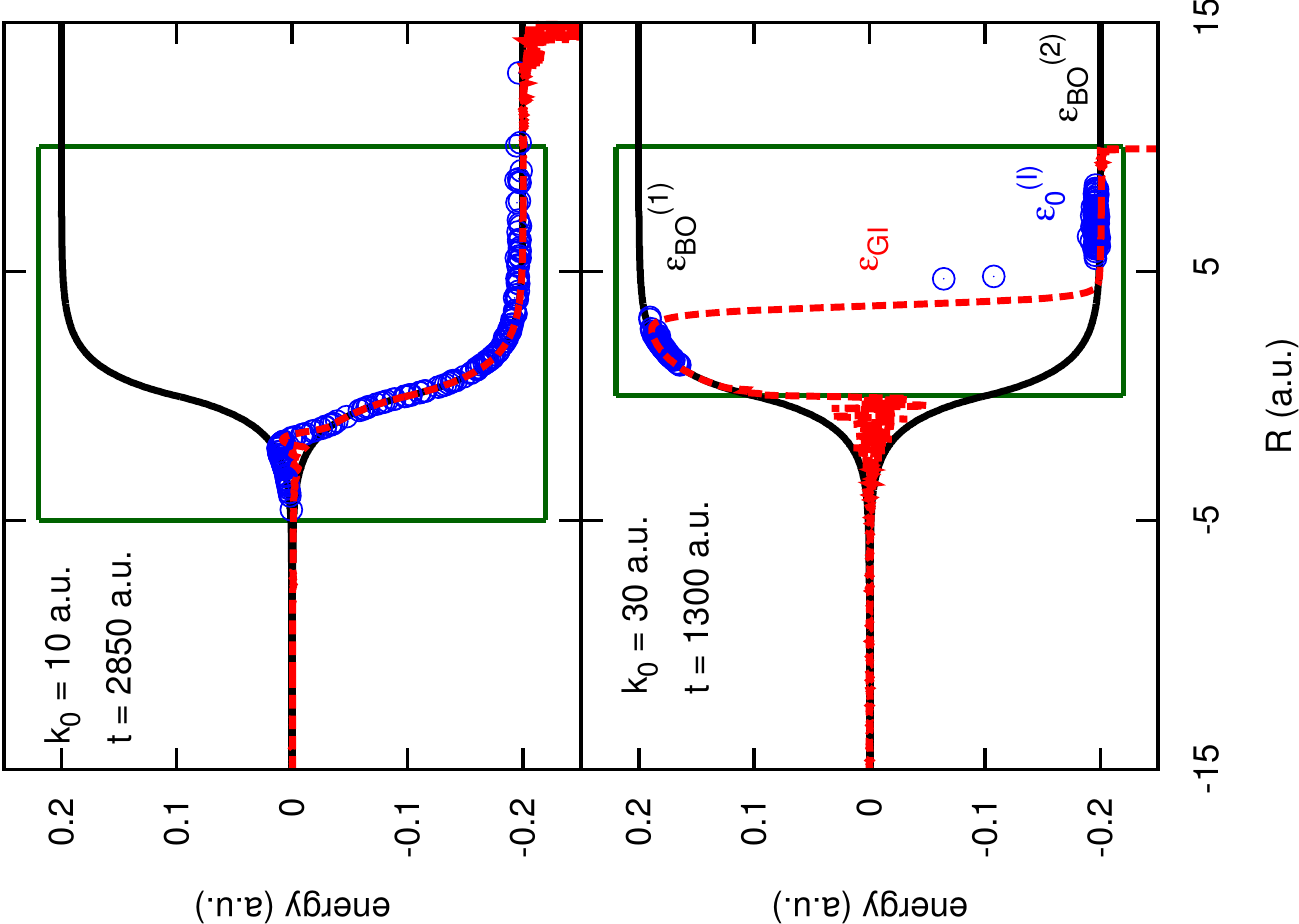}
  \includegraphics[width=0.5\textwidth,angle=270]{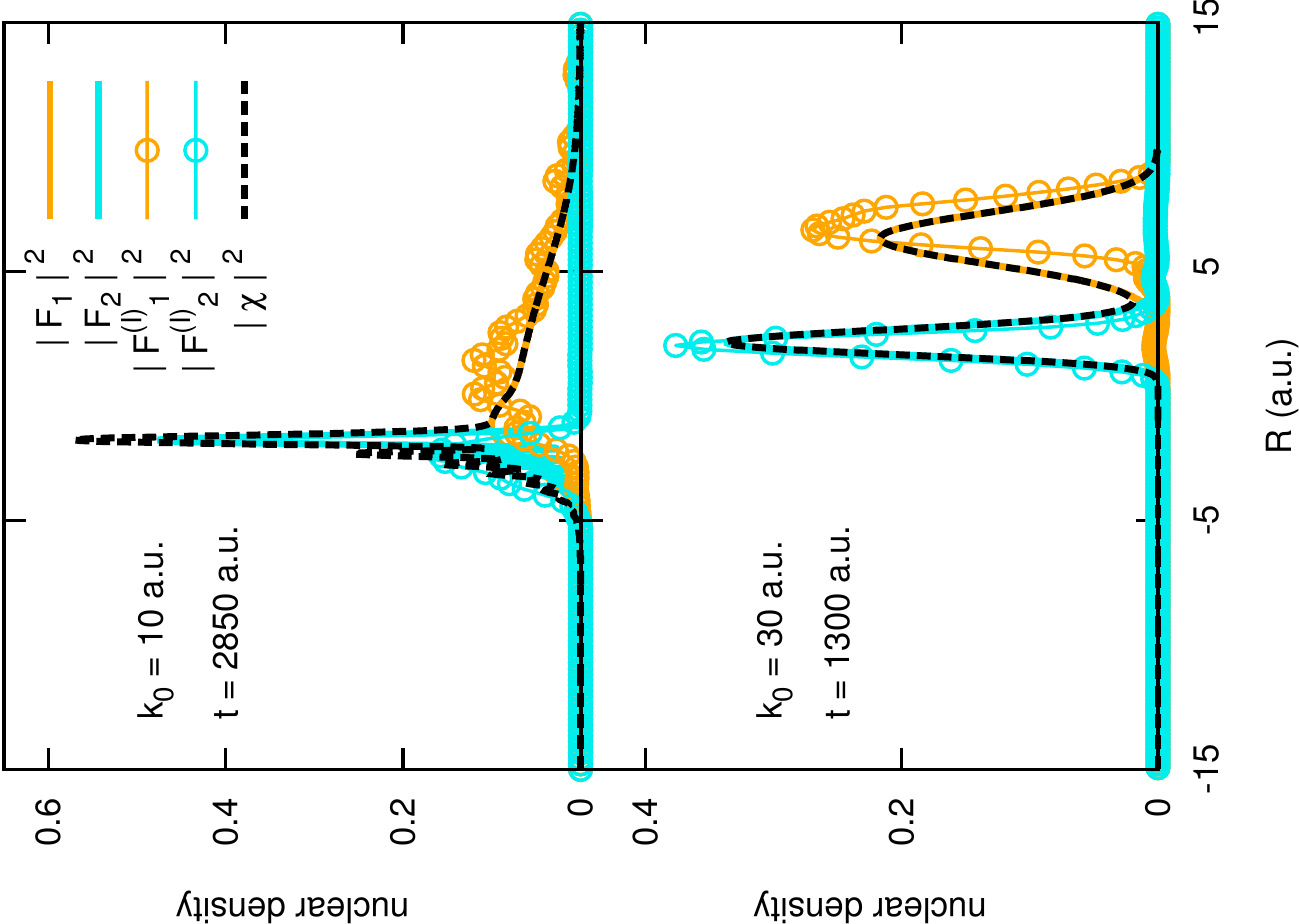}
 \end{center}
 \caption{Same as in Fig.~\ref{fig: model1b}, but the snapshots are shown at time $t=2850$~a.u. for the low initial momentum $\hbar k_0=10$~a.u. (upper panel) and at time $t=1300$~a.u. for the high initial momentum $\hbar k_0=30$~a.u. (lower panel), for model (c).}
 \label{fig: model3b}
\end{figure}

Figs.~\ref{fig: model3a} and~\ref{fig: model3b} show results for two values of the initial momentum, namely $\hbar k_0=10$~a.u. and $\hbar k_0=30$~a.u., for model system (c), in analogy to what has been presented in the previous examples. In particular, we notice that for the low initial momentum, the CT-MQC is able to reproduce the time-dependence of the adiabatic populations in very close agreement with exact results, yielding a better agreement than TSH. The population exchange occurring after 4000~a.u. (upper left panel of Fig.~\ref{fig: model3a}) is not captured by Ehrenfest and the MQC algorithm, since these approaches cannot reproduce nuclear dynamics along diverging paths, as already discussed in previous work~\cite{Gross_EPL2014, Gross_JCP2014}. Such second non-adiabatic event is observed when the reflected wave packet crosses for the second time the extended coupling region. This channel is not accessible in Ehrenfest and  MQC calculations.

The CT-MQC yields decoherence effects in strikingly good agreement with exact results. The electronic equations in TSH and Ehrenfest are given by the first two terms on the right-hand-side of Eq.~(\ref{eqn: final electronic eqn}) and, as shown in the lower panels in Fig.~\ref{fig: model3a}, those procedures completely miss the decay of coherence predicted by wave packet propagation. The additional term in Eq.~(\ref{eqn: final electronic eqn}) containing the quantum momentum is sufficient to correct for this effect.

As expected, also the TDPES and the nuclear densities are very well reproduced by the CT-MQC procedure, when compared to exact calculations.

\subsection{(d) - Double arch}
The double arch model has been introduced~\cite{subotnikJCP2011_1} to enhance decoherence effects on electronic dynamics. For low initial momenta, this model is similar to model (c), since the nuclear wave packet is partially transmitted and partially reflected. After the splitting though, both components go a second time through a region of extended coupling, thus they experience once again a non-adiabatic event. We observe this behavior when $\hbar k_0=20$~a.u. is chosen as initial momentum. By contrast, at higher initial momentum, i.e. $\hbar k_0=40$~a.u., both wave packets are transmitted and propagate in the positive region until they recombine. Capturing correctly the dynamics of the recombination is what makes this model system a challenge for the inclusion of decoherence effects. After the two transmitted wave packets propagate independently of each other on the two (very different) adiabatic surfaces, they recombine with some time-delay, therefore decoherence should first decay and then reappear as consequence of the recombination.

\begin{figure}[h!]
 \begin{center}
  \includegraphics[width=0.4\textwidth]{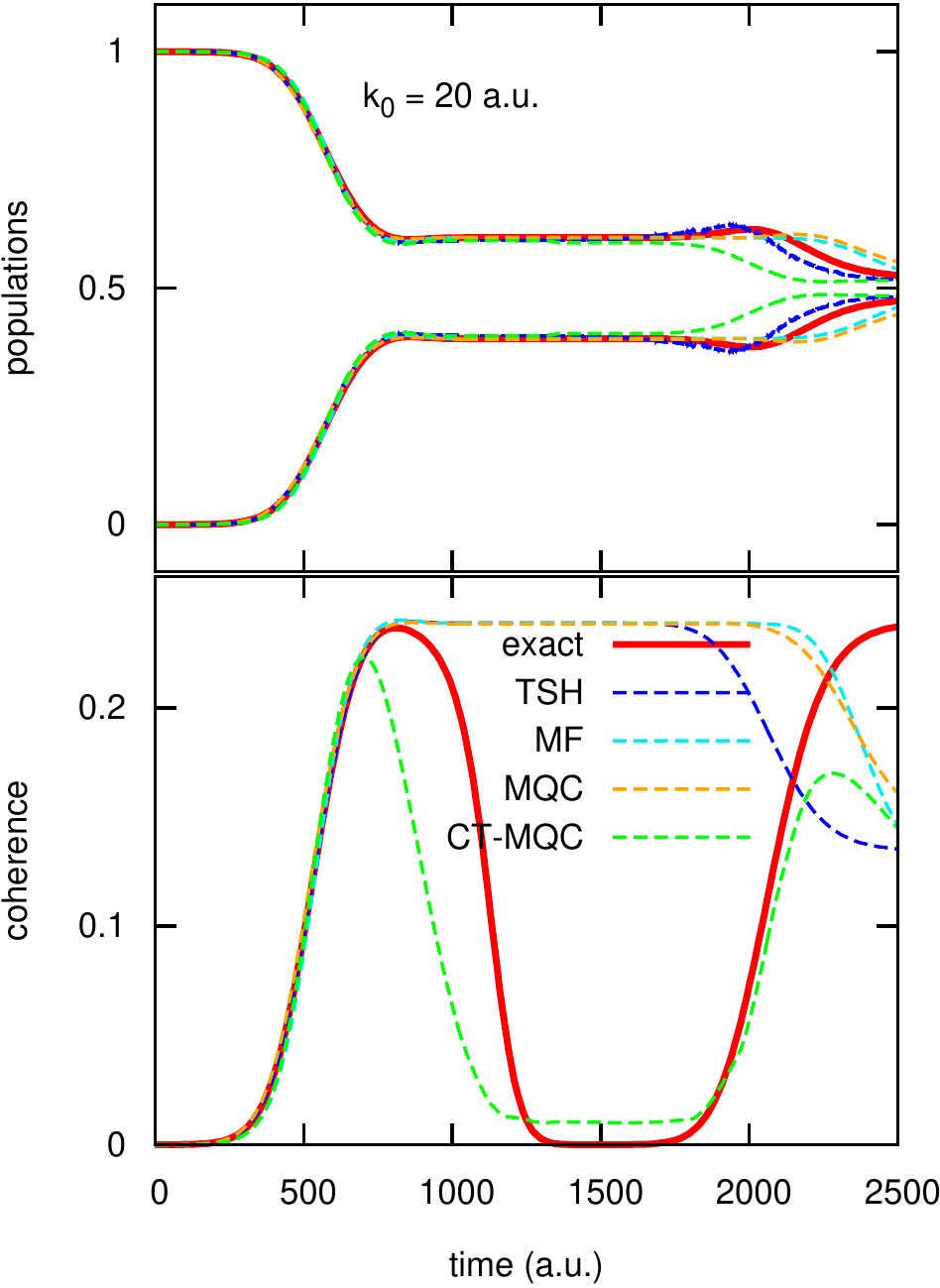}
  \includegraphics[width=0.4\textwidth]{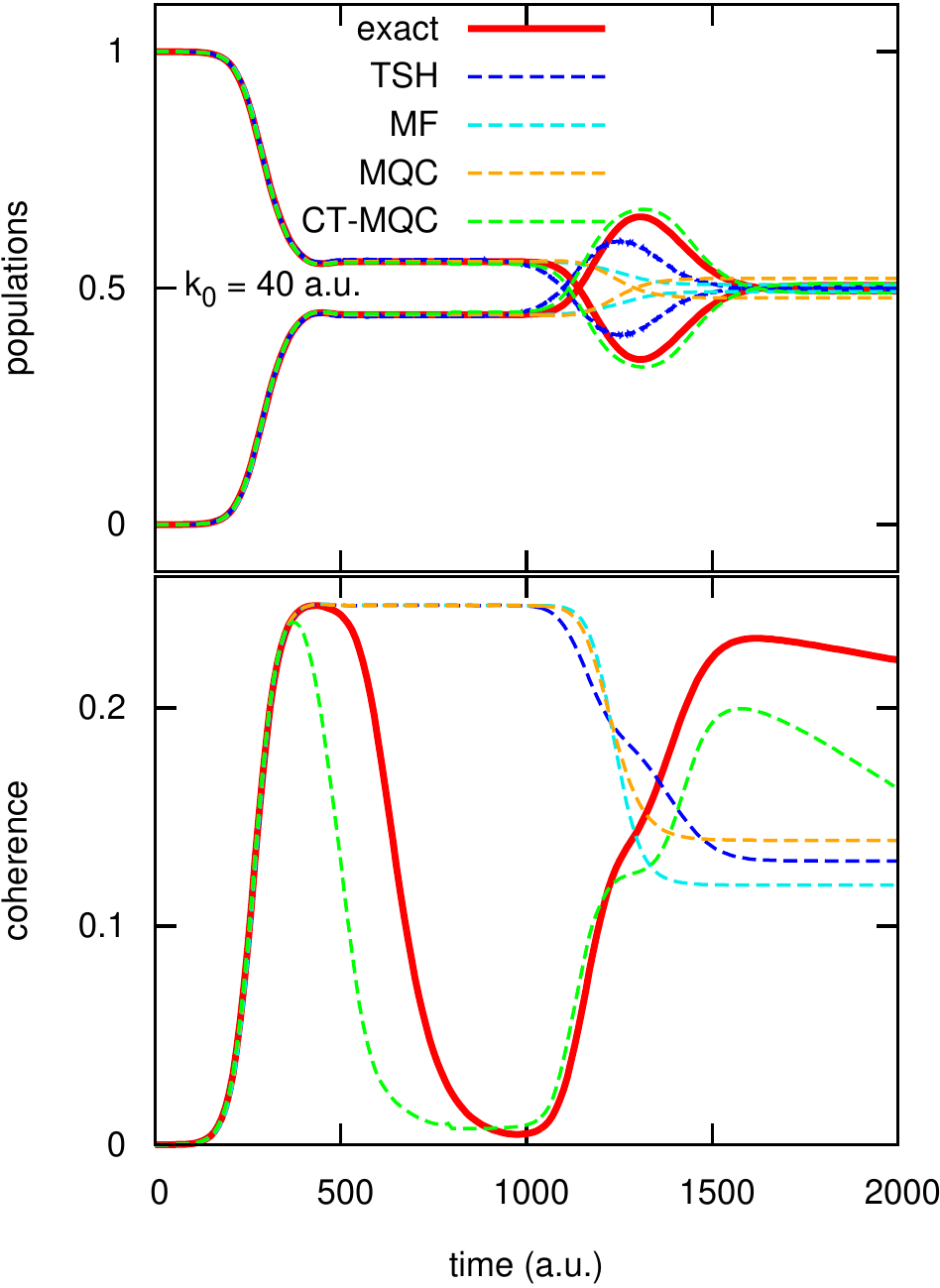}
 \end{center}
 \caption{(Left) Same as in Fig.~\ref{fig: model1a} but for the model system (d) and for the value $\hbar k_0=20$~a.u. of the initial momentum. (Right) Same as in the left panel but for the value $\hbar k_0=40$~a.u. of the initial momentum.}
 \label{fig: model4a}
\end{figure}

Fig.~\ref{fig: model4a} shows that all features related to decoherence discussed above are indeed captured by the CT-MQC algorithm, in very good agreement with exact results, and are completely missed by all other methods. At high initial momentum, only TSH and CT-MQC are able to reproduce the electronic population after the second non-adiabatic event after 1000~a.u. (as shown in Fig.~\ref{fig: model4a} upper right panel), but only CT-MQC results are in perfect agreement with exact results.

The non-adiabatic process represented in model (d), as well as in model (c), presents very different BO surfaces. Therefore, the correct nuclear dynamics cannot be captured by methods, such as Ehrenfest and MQC, relying on a single potential energy surface which is (or is close to) an average potential. Such an average potential cannot reproduce the very different forces that are experienced by the nuclear trajectories in different regions of space. By contrast, the single TDPES from the exact factorization is able to capture very different shapes because of the steps~\cite{Gross_PRL2013} that bridge piecewise adiabatic shapes. At a given time, depending on where the classical trajectory is located, it can experience very different forces.
\begin{figure}[h!]
 \begin{center}
  \includegraphics[width=0.5\textwidth,angle=270]{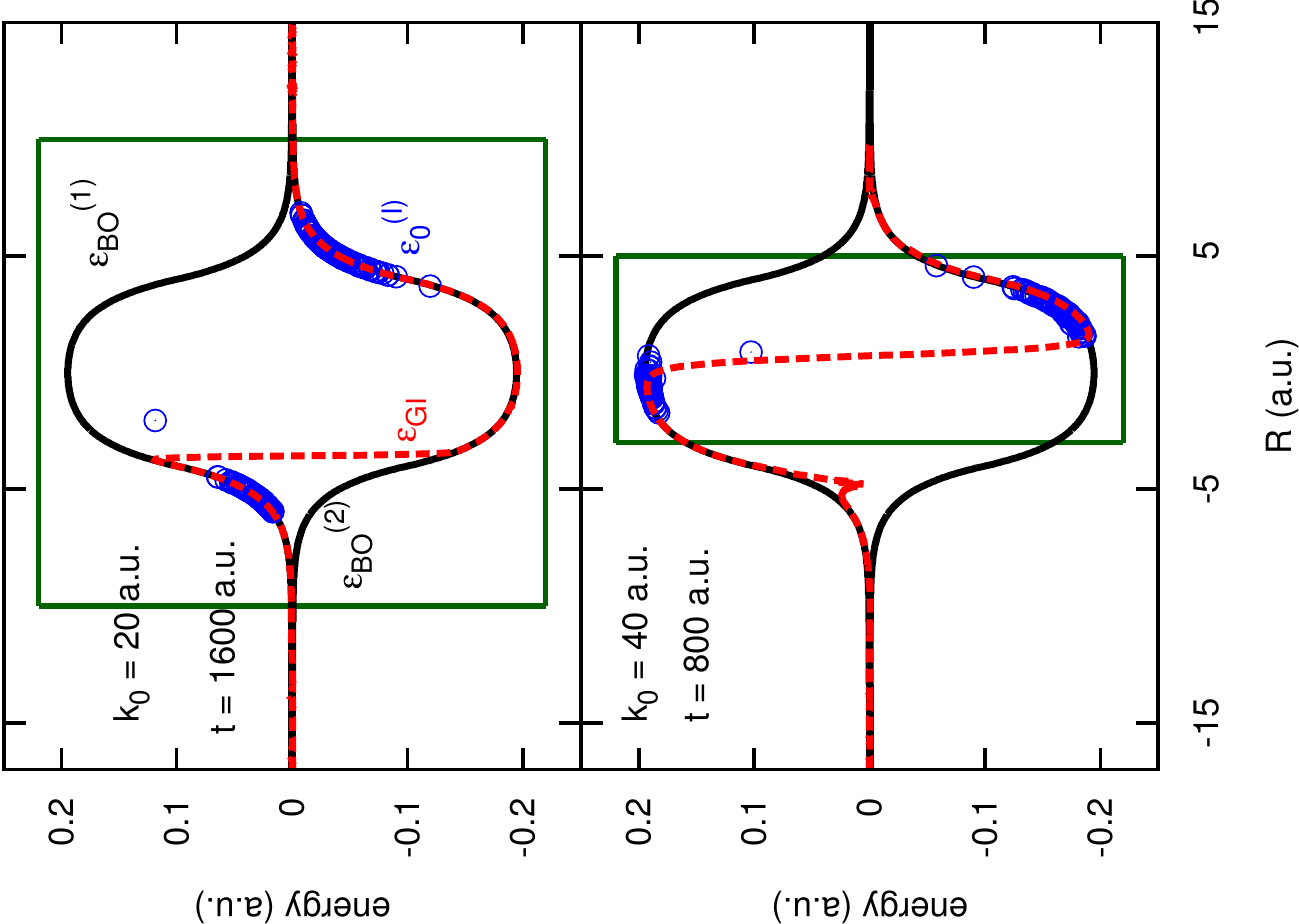}
  \includegraphics[width=0.5\textwidth,angle=270]{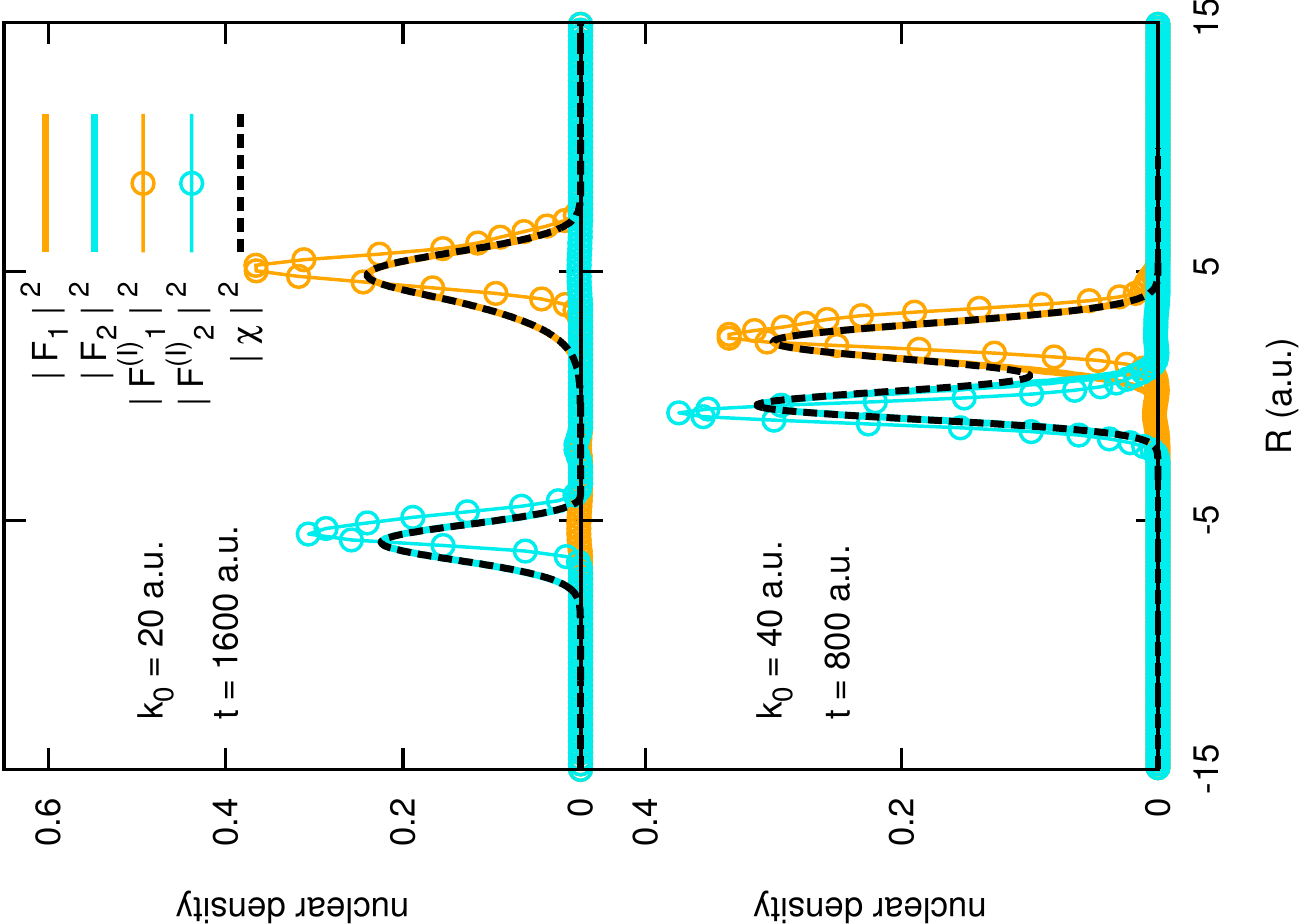}
 \end{center}
 \caption{(Left) Same as in Fig.~\ref{fig: model1b}, but the snapshots are shown at time $t=1600$~a.u. for the low initial momentum $\hbar k_0=20$~a.u. (upper panel) and at time $t=800$~a.u. for the high initial momentum $\hbar k_0=40$~a.u. (lower panel), for model (d).}
 \label{fig: model4b}
\end{figure}
This is clearly shown in Fig.~\ref{fig: model4b}.

\subsection{Dependence on the initial momentum}
It is interesting to investigate the dependence of the final transmission/reflection probabilities on the initial momentum of the nuclear nuclear wave packet for the four model systems studied here. Similar studies have been already reported in the literature~\cite{tully1990, subotnikJCP2011_1, shenvi-subotnikJCP2011, tavernelliJCP2013}, focussing on the performance of TSH in comparison to quantum propagation techniques. Here, we will only benchmark the CT-MQC algorithm against exact results. We refer to the above-mentioned references for TSH results, since we are considering the same initial conditions used there. Moreover, we will not show the results for Ehrenfest and for the independent-trajectory version of the MQC algorithm derived from the exact factorization. The reason is that neither method is capable of following the evolution of nuclear wave packets along diverging paths (at least with the initial conditions chosen below), as we have previously discussed~\cite{Gross_JCP2014}.

In all cases, the initial variance of the nuclear wave packet is chosen as $\sigma=20/(\hbar k_0)$. The initial conditions for the classical trajectories are sampled in position space from a Gaussian distribution with variance $\sigma$, while the same momentum, $\hbar k_0$, is assigned to all trajectories. For model (d), we have only chosen large values of $\hbar k_0$, since for lower values model (d) is similar to model (c).

Figure~\ref{fig: k0 studies, model a to d} shows a good agreement between exact and quantum-classical results for models (a), (c) and (d). As expected from the results reported for model (b) in Fig.~\ref{fig: model2a} at low initial momentum, the CT-MQC is not able to reproduce the final probabilities up to $\hbar k_0=20$~a.u., but it improves at higher values. Still, it slightly underestimates the non-adiabatic population exchange up to a value of $\hbar k_0=26$~a.u. of the initial momentum.
\begin{figure}[h!]
 \begin{center}
  \includegraphics[width=0.6\textwidth,angle=270]{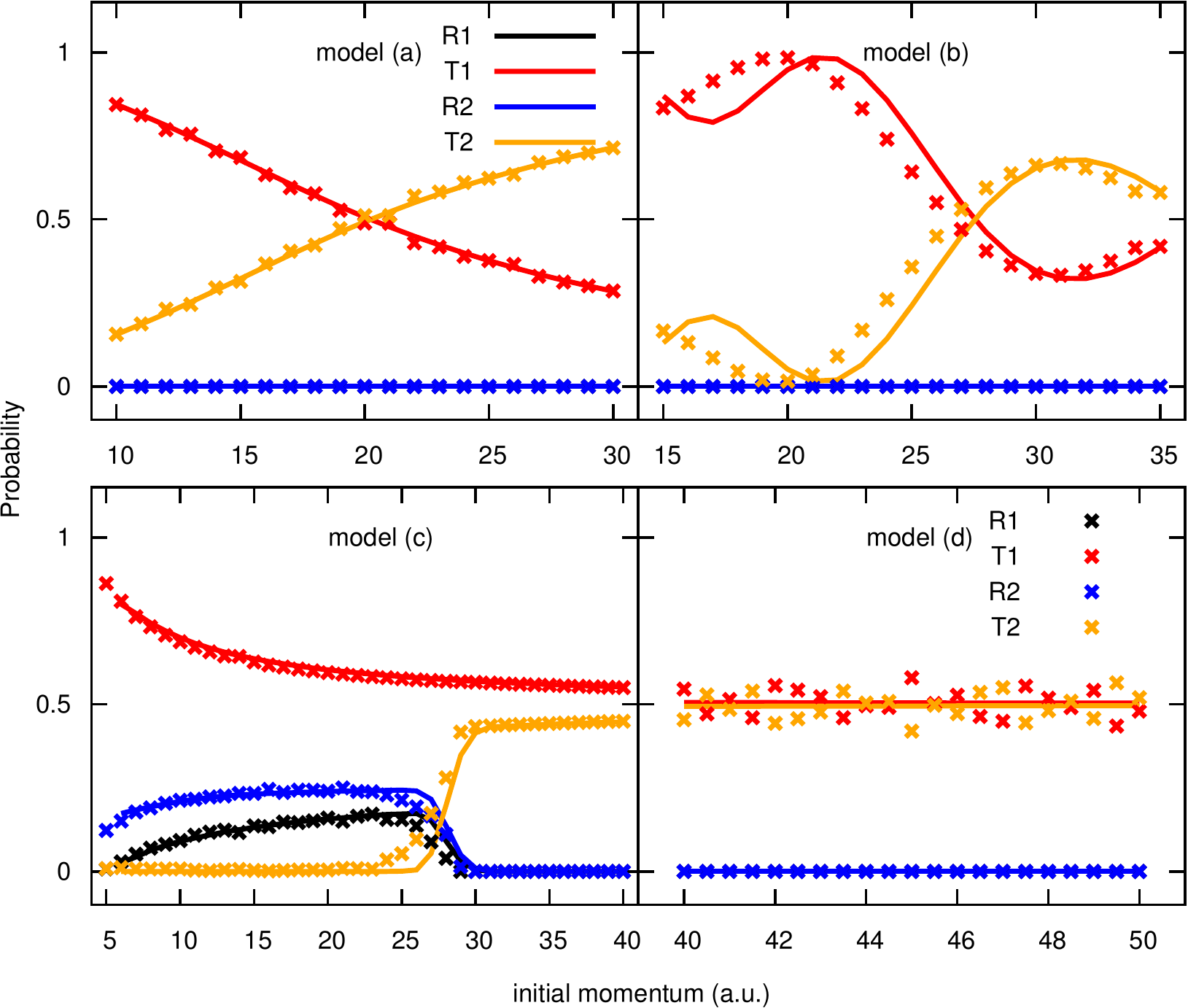}
 \end{center}
 \caption{Dependence of the final transmission/reflection probabilities on the initial momentum $\hbar k_0$. Transmission probabilities at the final times on the lower and upper surfaces are indicated as T1 and T2, respectively; similarly, the reflection probabilities are indicated as R1 and R2. The lines refer to exact calculations, the dots to the results of the CT-MQC algorithm. The initial variance of the nuclear wave packet is chosen as $\sigma=20/(\hbar k_0)$.}
 \label{fig: k0 studies, model a to d}
\end{figure}
Also, it is worth noting that the (well-known) unphysical oscillations in model (c), observed for the reflection channels in TSH~\cite{tully1990, subotnikJCP2011_1}, are not present in our CT-MQC results. In TSH, the oscillations appear when the initial momenta are not sampled from a distribution~\cite{shenviJCP2011_3} but a fixed value of $\hbar k_0$ is assigned to all trajectories, as done here. They are due to the phase coherence between the first and the second passage of the wave packet through the region of extended coupling. The CT-MQC algorithm correctly eliminates any sign of spurious coherence, that is in agreement with quantum mechanical results.

In model (d), the final probabilities shown in Fig.~\ref{fig: k0 studies, model a to d} slightly oscillate around the exact value, but the deviation is not larger than 10\%. The model has been initially introduced~\cite{subotnikJCP2011_1} to enhance the over-coherence problem of TSH and to investigate possible correction strategies. Some studies have been reported~\cite{subotnikJCP2011_1, tavernelliJCP2013} where the initial width of the nuclear wave packet is larger than the cases above, i.e. $\sigma=100/(\hbar k_0)$, which yields a very localized wave packet in momentum space. We have simulated also this situation, following the same protocol as described above for sampling classical initial conditions, for the same range of $\hbar k_0$. In a recent work~\cite{tavernelliJCP2013}, Curchod and Tavernelli have shown that TSH (i) is not able to reproduce the oscillations in the transmitted probabilities, if the initial positions and momenta of the classical trajectories are sampled from Gaussian distributions, and (ii) captures the oscillatory behavior of the transmitted probabilities (even though not the correct one), if the same positions and momenta are chosen for all trajectories, which will then differ from each other at later times, due to the stochastic jumps. In the CT-MQC algorithm, if option (ii) is chosen, all trajectories will follow the same path and no splitting will be observed. This is intrinsic of the deterministic nature of the algorithm. Therefore, we chose the option to sample the initial positions from a Gaussian distribution and we assigned the same initial momentum $\hbar k_0$ to all trajectories. As it can be seen in Fig.~\ref{fig: k0 studies, model d}, the CT-MQC algorithm is not able to match the exact results.
\begin{figure}[h!]
 \begin{center}
  \includegraphics[width=0.4\textwidth,angle=270]{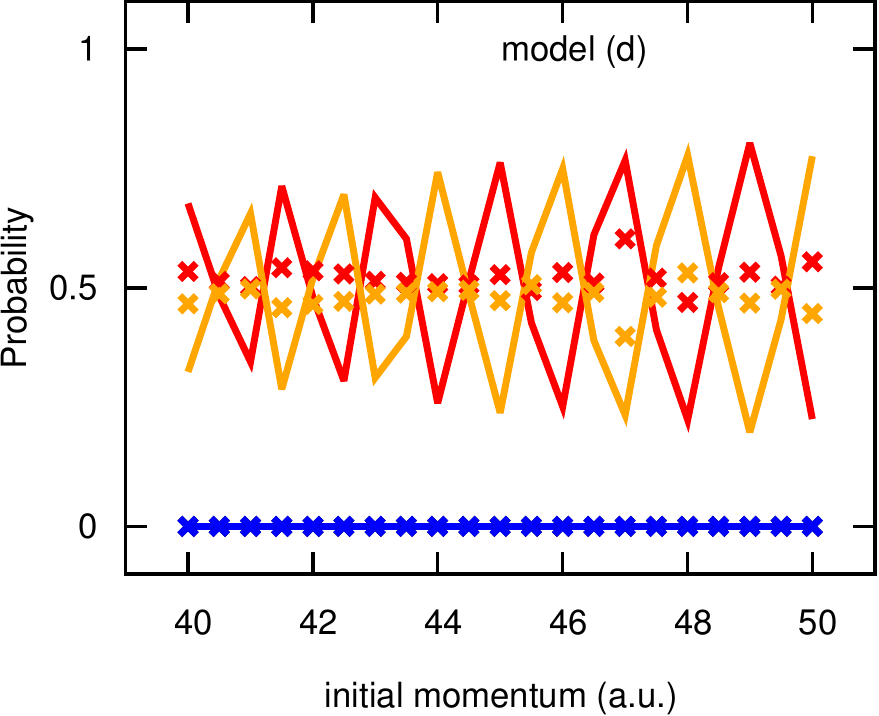}
 \end{center}
 \caption{Dependence of the final transmission/reflection probabilities on the initial momentum $\hbar k_0$ for model (d). The initial variance of the nuclear wave packet is chosen as $\sigma=100/(\hbar k_0)$.}
 \label{fig: k0 studies, model d}
\end{figure}
Here, we would like to emphasise that this a very pathological situation, introduced to exaggerate the over-coherence issue of TSH. Further studies and more accurate approximations might be necessary to cure this problem in CT-MQC.

\section{Conclusions}\label{sec: conclusions}
We have presented a detailed derivation of the CT-MQC scheme~\cite{Gross_PRL2015} to non-adiabatic dynamics starting from the exact factorization of the electron-nuclear wave function. When compared to the independent-trajectory MQC~\cite{Gross_EPL2014, Gross_JCP2014}, the CT-MQC offers a clear improvement, since the new method is now able to correctly reproduce electronic decoherence and the spatial splitting of the nuclear wave packet, which in turn is connected to the ability of reproducing the steps~\cite{Gross_PRL2013,Gross_MP2013,Gross_JCP2015} in the gauge-invariant part of the TDPES. In the proposed scheme, the classical trajectories are indirectly coupled through the electronic equation. This feature is therefore different from the direct interaction among trajectories provided by the quantum potential in Bohmian dynamics.

Two key features have been included here, that go beyond our previous lowest order algorithm, (i) the quantum momentum appears in the electronic equation, when considering $\mathcal O(\hbar)$ terms in the expression of the ENCO depending on the nuclear wave function, and (ii) the spatial dependence of the coefficients in the Born-Huang expansion of the electronic wave function has been taken into account to leading order, while it was completely neglected in our previous work. In the new picture we have been able to distinguish the two effects induced by nuclear motion on electronic dynamics, populations dynamics and decoherence. On the one hand, non-adiabatic transitions are induced by the nuclear momentum, the zero-th order term of the $\hbar$-expansion, which couples to the NACVs. On the other hand, decoherence effects are controlled by the quantum momentum, the first-order term of the $\hbar$-expansion, which is an imaginary correction to the nuclear momentum in the ENCO. In turn, the effect of the electrons on nuclear dynamics is represented by the TDPES and the TDVP. As discussed in previous work~\cite{Gross_PRL2013, Gross_MP2013, Gross_JCP2015}, being able to reproduce their features, such as the steps, in an approximate scheme results in the correct description of the nuclear wave packet behavior. 

In this paper the analytical development of the algorithm is supported by tests performed on some model systems~\cite{tully1990, subotnikJCP2011_1} which are typical examples of electronic non-adiabatic processes. The comparison of the CT-MQC results with wave packet dynamics has shown that indeed we are able to predict the correct physical picture, capturing the dynamical details of both electronic and nuclear subsystems. Moreover, we have proved that the new algorithm, being a coupled-trajectory scheme, is able to overcome the issues that are encountered when other methods, based on an independent-trajectory picture, are employed.

Further developments are envisaged, mainly focusing on the application of the method to more realistic problems and on the semiclassical treatment of nuclear dynamics, possibly allowing the treatment of nuclear interference effects.

\section*{Acknowledgements}
Partial support from the Deutsche Forschungsgemeinschaft (SFB 762) and from the European Commission (FP7-NMP-CRONOS) is gratefully acknowledged. This work was supported by the 2015 Research Fund (1.150059.01) of UNIST (Ulsan National Institute of Science \& Technology)
A. A. gratefully acknowledges the support by the European Research Council Advanced Grant DYNamo (ERC- 2010-AdG-267374) and Grupo Consolidado UPV/EHU del Gobierno Vasco (IT578-13).

\section*{Supporting Information Available}
Section SI.1 of the Supporting Information introduces the exact factorization framework, recalling the general theory and the equations that are the starting point for the approximations developed in the main text.

Section SI.2 justifies the approximations leading to Eq.~(\ref{eqn: nabla gamma}) by using semiclassical arguments. Moreover, we recall there the analysis of the coefficients $|C_l(\dulR,t)|$ of the Born-Huang expansion of the electronic wave function that we presented in previous work~\cite{Gross_PRL2013}.

The equations to compute the quantum momentum in a multi-level system are derived in Section SI.3 of the Supporting Information and are presented as a generalization of the two-level case discussed in the main text.

Finally, Section SI.4 gives the computational details for the numerical tests.

This information is available free of charge via the Internet at http://pubs.acs.org/.

\addcontentsline{toc}{section}{References}
%\bibliography{./biblio,./factorization,./mqc}

\section*{Table of Contents graphic}

\begin{figure}%[h!]
 \begin{center}
  \includegraphics[width=0.9\textwidth]{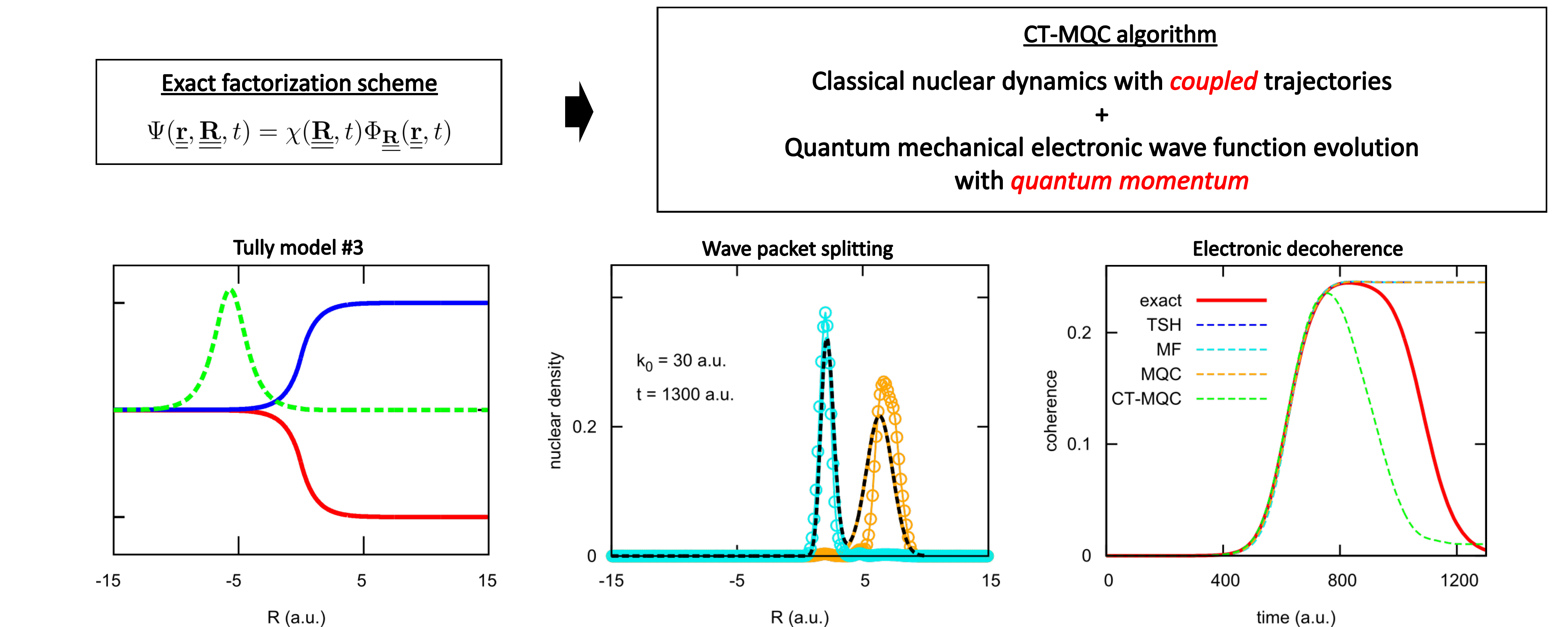}
 \end{center}
\end{figure}

\end{document}

\title{Supporting Information to ``Quantum-Classical Non-Adiabatic Dynamics: Coupled- vs. Independent-Trajectory Methods''}

\author{Federica Agostini}
\affiliation{Max-Planck Institut f\"ur Mikrostrukturphysik, Weinberg 2, D-06120 Halle, Germany}
\author{Seung Kyu Min}
\affiliation{Ulsan National Institute of Technology}
\affiliation{Max-Planck-Institut f\"ur Mikrostrukturphysik, Weinberg 2, D-06120 Halle, Germany}
\email{skmin@unist.ac.kr}
\author{Ali Abedi}
\affiliation{Nano-Bio Spectroscopy group, Dpto. F{\'i}sica de Materiales, Universidad del Pa{\'i}s Vasco,
Centro de F{\'i}sica de Materiales CSIC-UPV/EHU-MPC and DIPC, Av. Tolosa 72, E-20018 San Sebasti{\'a}n, Spain}
\author{E. K. U. Gross}
\affiliation{Max-Planck-Institut f\"ur Mikrostrukturphysik, Weinberg 2, D-06120 Halle, Germany}

\maketitle

\section{SI.1 - Exact factorization of the electron-nuclear wave function}\label{sec: background}
The non-relativistic Hamiltonian describing a system of interacting electrons and nuclei, in the absence of a time-dependent external field, is
\begin{equation}\label{eqn: hamiltonian}
 \hat H = \hat T_n+\hat H_{BO},
\end{equation}
where $\hat T_n$ is the nuclear kinetic energy operator and 
\begin{equation}\label{eqn: boe}
\hat{H}_{BO}(\dulr,\dulR) = \hat{T}_e(\dulr) + \hat{W}_{ee}(\dulr) + \hat{V}_{en}(\dulr,\dulR) +
\hat{W}_{nn}(\dulR)
\end{equation}
is the standard BO electronic Hamiltonian, with electronic kinetic energy $\hat{T}_e(\dulr)$, and interaction potentials  $\hat{W}_{ee}(\dulr)$ for electron-electron, $\hat{W}_{nn}(\dulR)$ for nucleus-nucleus, and  $\hat{V}_{en}(\dulr,\dulR)$ for electron-nucleus. The symbols $\dulr$ and $\dulR$ are used to collectively indicate the coordinates of $N_{e}$ electrons and $N_n$ nuclei, respectively.

It has been proved~\cite{Gross_PRL2010,Gross_JCP2012}, that the full time-dependent electron-nuclear wave function, $\Psi(\dulr,\dulR,t)$, that is the solution of the time-dependent Schr\"odinger equation (TDSE),
\begin{equation}\label{eqn: tdse}
 \hat H\Psi(\dulr,\dulR,t)=i\hbar\partial_t \Psi(\dulr,\dulR,t),
\end{equation}
can be exactly factorized to the product
\begin{equation}\label{eqn: factorization}
 \Psi(\dulr,\dulR,t)=\chi(\dulR,t)\Phi_\dulR(\dulr,t)
\end{equation}
 where 
\begin{equation}
 \int d\dulr \left|\Phi_\dulR(\dulr,t)\right|^2 = 1 \quad\forall\,\,\dulR,t.
\label{eq:PNC}
\end{equation}
Here $\chi(\dulR,t)$ is the nuclear wave function and $\Phi_\dulR(\dulr,t)$ is the electronic wave function  which parametrically depends on the nuclear positions and satisfies the partial normalization condition (PNC) expressed in Eq.~(\ref{eq:PNC}). The PNC guarantees the interpretation of $|\chi(\dulR,t)|^2$ as the probability of finding the nuclear configuration $\dulR$ at time $t$, and of $|\Phi_\dulR(\dulr,t)|^2$ itself as the conditional probability of finding the electronic configuration $\dulr$ at time $t$ for nuclear configuration $\dulR$.  Further, the PNC makes the factorization~(\ref{eqn: factorization}) unique up to within a $(\dulR,t)$-dependent gauge transformation,
\begin{equation}\label{eqn: gauge}
 \begin{array}{rcl}
  \chi(\dulR,t)\rightarrow\tilde\chi(\dulR,t)&=&e^{-\frac{i}{\hbar}\theta(\dulR,t)}\chi(\dulR,t) \\
  \Phi_\dulR(\dulr,t)\rightarrow\tilde\Phi_\dulR(\dulr,t)&=&e^{\frac{i}{\hbar}\theta(\dulR,t)}\Phi_\dulR(\dulr,t),
 \end{array}
\end{equation}
 where $\theta(\dulR,t)$ is some real function of the nuclear coordinates and time. 

The stationary variations~\cite{frenkel} of the quantum mechanical action with respect to $\Phi_\dulR(\dulr,t)$ and $\chi(\dulR,t)$ lead to the derivation of the following equations of motion
\begin{eqnarray}
 \left(\hat H_{el}(\dulr,\dulR)-\epsilon(\dulR,t)\right)
 \Phi_{\dulR}(\dulr,t)&=&i\hbar\partial_t \Phi_{\dulR}(\dulr,t)\label{eqn: exact electronic eqn} \\ 
 \hat H_n(\dulR,t)\chi(\dulR,t)&=&i\hbar\partial_t \chi(\dulR,t), \label{eqn: exact nuclear eqn}
\end{eqnarray}
where the PNC is enforced in the variational principle by means of Lagrange multipliers~\cite{alonsoJCP2013,Gross_JCP2013}. Here, the electronic and nuclear Hamiltonians are defined as
\begin{equation}\label{eqn: electronic hamiltonian}
\hat H_{el}(\dulr,\dulR)=\hat{H}_{BO}(\dulr,\dulR)+\hat U_{en}[\Phi_\dulR,\chi]
\end{equation}
and
\begin{equation}\label{eqn: nuclear hamiltonian}
\hat H_n(\dulR,t) = \sum_{\nu=1}^{N_n} \frac{\left[-i\hbar\nabla_\nu+\bA_\nu(\dulR,t)\right]^2}{2M_\nu} + \epsilon(\dulR,t),
\end{equation}
respectively, with the ``electron-nuclear coupling'' operator (ENCO) 
\begin{align}
\hat U_{en}[\Phi_\dulR,\chi]=&\sum_{\nu=1}^{N_n}\frac{1}{M_\nu}\left[
 \frac{\left[-i\hbar\nabla_\nu-\bA_\nu(\dulR,t)\right]^2}{2}+\left(\frac{-i\hbar\nabla_\nu\chi}{\chi}+\bA_\nu(\dulR,t)\right)
 \left(-i\hbar\nabla_\nu-\bA_{\nu}(\dulR,t)\right)\right].\label{eqn: enco}
\end{align}
The exact potentials appearing in the theory are scalar time-dependent potential energy surface (TDPES), $\epsilon(\dulR,t)$, implicitly defined by Eq.~(\ref{eqn: exact electronic eqn}) as
\begin{align}
 \epsilon(\dulR,t)=\left\langle\Phi_\dulR(t)\right|\hat{H}_{BO}+\hat U_{en}-i\hbar\partial_t\left|
 \Phi_\dulR(t)\right\rangle_\dulr=\epsilon_{GI}\left(\dulR,t\right)+\left\langle\Phi_\dulR(t)\right|-i\hbar\partial_t\left|
 \Phi_\dulR(t)\right\rangle_\dulr, \label{eqn: tdpes}
\end{align}
with $\epsilon_{GI}(\dulR,t)$ indicating the gauge-invariant (GI) part of the TDPES, and the time-dependent vector potential, $\bA_{\nu}\left(\dulR,t\right)$, defined as
\begin{equation}\label{eqn: vector potential}
 \bA_{\nu}\left(\dulR,t\right) = \left\langle\Phi_\dulR(t)\right|-i\hbar\nabla_\nu\left.\Phi_\dulR(t)
 \right\rangle_\dulr\,.
\end{equation}
The symbol $\left\langle\,\,\cdot\,\,\right\rangle_\dulr$ indicates an integration over electronic coordinates only. Under the gauge transformation~(\ref{eqn: gauge}), the scalar potential and the vector potential transform as  
\begin{eqnarray}
\tilde{\epsilon}(\dulR,t) &=& \epsilon(\dulR,t)+\partial_t\theta(\dulR,t)\label{eqn: transformation of epsilon} \\
\tilde{\bf A}_{\nu}(\dulR,t) &=& {\bf A}_{\nu}(\dulR,t)+\nabla_\nu\theta(\dulR,t)\,.\label{eqn: transformation of A}
\end{eqnarray}
In Eqs.~(\ref{eqn: exact electronic eqn}) and~(\ref{eqn: exact nuclear eqn}), $\hat U_{en}[\Phi_\dulR,\chi]$, $\epsilon(\dulR,t)$ and $\bA_{\nu}\left(\dulR,t\right)$ are responsible for the coupling between electrons and nuclei in a formally exact way. It is worth noting that the electron-nuclear coupling operator, $\hat U_{en}[\Phi_\dulR,\chi]$, in the electronic equation~(\ref{eqn: exact electronic eqn}), depends on the nuclear wave function and acts on the parametric dependence of $\Phi_\dulR(\dulr,t)$ as a differential operator. This ``pseudo-operator'' includes the coupling to the nuclear subsystem beyond the parametric dependence in the BO Hamiltonian $\hat H_{BO}(\dulr,\dulR)$.

The nuclear equation~(\ref{eqn: exact nuclear eqn}) has the particularly appealing form of a Schr\"odinger equation that contains a time-dependent scalar potential~(\ref{eqn: tdpes}) and a time-dependent vector potential~(\ref{eqn: vector potential}) that govern the nuclear dynamics and yield the nuclear wave function. The scalar and vector potentials are uniquely determined up to within a gauge transformation, given in Eqs.~(\ref{eqn: transformation of epsilon}) and~(\ref{eqn: transformation of A}). As expected, the nuclear Hamiltonian in Eq.~(\ref{eqn: exact nuclear eqn}) is form-invariant under such transformations. $\chi(\dulR,t)$ is interpreted as the nuclear wave function since it leads to an $N$-body nuclear density,
\begin{equation}
\Gamma(\dulR,t)=\vert\chi(\dulR,t)\vert^2,
\end{equation}
 and an $N$-body current density, 
\begin{equation} 
 {\bf J}_\nu(\dulR,t)=\frac{\Big[\mbox{Im}(\chi^*(\dulR,t)\nabla_\nu\chi(\dulR,t))+ \Gamma(\dulR,t){\bf A}_\nu(\dulR,t)\Big]}{M_\nu},
\end{equation}
which reproduce the true nuclear $N$-body density and current density~\cite{Gross_JCP2012} obtained from the full wave function $\Psi(\dulr,\dulR,t)$. The uniqueness of $\epsilon(\dulR,t)$ and $\bA_{\nu}(\dulR,t)$ (up to within the gauge transformation~(\ref{eqn: transformation of epsilon}) and~(\ref{eqn: transformation of A})) can be straightforwardly proved by following the steps of the current density version~\cite{Ghosh-Dhara} of the Runge-Gross theorem~\cite{RGT}, or by referring to the theorems proved in Ref.~\cite{Gross_PRL2010}. 

As in previous work~\cite{Gross_JCP2012, Gross_PRL2013, Gross_MP2013, Gross_JCP2014, Gross_EPL2014, Gross_JCP2015, Gross_PRL2015} we will represent the electronic equation, and all quantities depending on the electronic wave function, on the adiabatic basis. Therefore, we introduce here the Born-Oppenheimer (BO) electronic states, $\varphi_{\dulR}^{(l)}(\dulr)$, and BO potential energy surfaces (PESs), $\epsilon_{BO}^{(l)}(\dulR)$, which are the normalized eigenstates and eigenvalues of the BO electronic Hamiltonian~(\ref{eqn: boe}), respectively. If the full wave function is expanded in this basis,
\begin{equation}\label{eqn: expansion of Psi}
 \Psi(\dulr,\dulR,t)=\sum_l F_l(\dulR,t)\varphi_\dulR^{(l)}(\dulr),
\end{equation} 
then the nuclear density may be written as  
\begin{equation}\label{eqn: chi and Fl}
 \left|\chi(\dulR,t)\right|^2 = \sum_{l}\left|F_l(\dulR,t)\right|^2.
\end{equation}
This relation is obtained by integrating the squared modulus of Eq.~(\ref{eqn: expansion of Psi}) over the electronic coordinates. The exact electronic wave function may also be expanded in terms of the BO states,
\begin{equation}\label{eqn: expansion of Phi}
 \Phi_\dulR(\dulr,t)=\sum_l C_l(\dulR,t)\varphi_\dulR^{(l)}(\dulr).
\end{equation} 
The expansion coefficients in Eqs.~(\ref{eqn: expansion of Psi}) and~(\ref{eqn: expansion of Phi}) are related,
\begin{equation}\label{eqn: relation coefficients}
 F_l(\dulR,t)= C_l(\dulR,t)\chi(\dulR,t),
\end{equation}
by virtue of the factorization~(\ref{eqn: factorization}). The PNC then reads
\begin{equation}\label{eqn: PNC on BO}
 \sum_l\left|C_l(\dulR,t)\right|^2=1\quad\forall\,\,\dulR,t.
\end{equation}

\section{SI.2 - Semiclassical analysis of the phases of the coefficients in the Born-Huang expansion}
Eq.~(16) is the final result of the \textit{third approximation} discussed in the paper and is related to the way the spatial derivatives with respect to the nuclear coordinates of the coefficients in the Born-Huang expansion~(\ref{eqn: expansion of Phi}) are estimated. 

First of all, we recall the discussion explaining the reason why the term $\nabla_\nu |C_l^{(I)}(t)|$ is neglected. Based on the solution of the full TDSE~\cite{Gross_PRL2013, Gross_MP2013, Gross_JCP2015} for one-dimensional systems, we have observed that at a given time the quantities $|C_l(\dulR,t)|$ are either constant functions of $\dulR$ or present a sigmoid shape~\cite{Gross_JCP2015}. In particular, this second feature appears when the nuclear wave density splits after having crossed a region of strong coupling. When they are constant, their spatial derivatives are zero, thus our approximation holds perfectly; in the sigmoid case, $|C_l(\dulR,t)|$ is constant, i.e. either 0 or 1, far from the step of the sigmoid function, and linear around the center of the step. Furthermore, the center of the step is the position where the nuclear density splits~\cite{Gross_PRL2013}, and only in this region the gradient $|C_l(\dulR,t)|$ might be significantly different from zero. Notice that these properties are very general, as we have analytically derived~\cite{Gross_JCP2015}, in the absence of an external time-dependent field. It is important to keep in mind that the nuclear density is reconstructed from the distribution of classical trajectories, thus it seems reasonable to assume that when the nuclear density splits into two or more branches, the probability of finding trajectories in the tail regions is very small. It follows that we are making the approximation $\nabla_\nu |C_l^{(I)}|/|C_l^{(I)}|=0$ only for few trajectories located in the tail regions. We expect that such an approximation will not affect dramatically the final (averaged over all trajectories) results. This statement is indeed confirmed by the numerical tests presented in the paper. At the initial times of the splitting event, trajectories can indeed be found in the region around the step because the nuclear density is not small. In such a region we then have~\cite{Gross_PRL2013,Gross_MP2013}
\begin{align}
 \frac{\partial_R\left|C_l^{(I)}(t)\right|}{\left|C_l^{(I)}(t)\right|} =
 \frac{\pm\alpha}{1/2\pm\alpha\left(R^{(I)}(t)-R_0\right)},
\end{align}
using a notation valid for one-dimensional cases, but easily applicable to higher dimensionality. The parameters $\alpha$ and $R_0$ characterize the step, i.e. they are its slope and center. The trajectories found in the vicinity of $R_0$, for which the relation $|R^{(I)}(t)-R_0|\sim 0$ is valid, will contribute to the spatial derivative of the coefficients as  $\partial_R|C_l^{(I)}|/|C_l^{(I)}|\sim \pm2\alpha$. This is the quantity that will be neglected. In comparison to $\mathbf f_{l,\nu}^{(I)}(t)$ in Eq.~(16), however, it has a finite value, while $\mathbf f_{l,\nu}^{(I)}(t)$ is given by the force produced by the $l$-th BO surface accumulated over time (it contains a time integral). It is worth noting that neglecting the term $\nabla_\nu|C_l^{(I)}|/|C_l^{(I)}|$ in the expression for the spatial derivative of $C_l^{(I)}$ does not affect the equation describing the evolution of the population of the BO states. This can be easily proved using Eq.~(14) to compute the time derivative of $|C_l^{(I)}(t)|^2$.

In order to derive Eq.~(16), we recall the factorization, in the form~(\ref{eqn: relation coefficients}), and we get the (exact) relation for the 
phases of $C_l$, $F_l$ and $\chi$,
\begin{align}
 \gamma_l\left(\dulR,t\right) = s_l\left(\dulR,t\right)-S\left(\dulR,t\right).
\end{align}
Here we indicate with the symbol $s_l\left(\dulR,t\right)$ and $S\left(\dulR,t\right)$ the phases of $F_l(\dulR,t)$ and of $\chi(\dulR,t)$, respectively. We interpret this relation semiclassically, by introducing the momenta $\nabla_\nu s_l\left(\dulR,t\right)=\mathbf p_{l,\nu}$ and $\nabla_\nu S\left(\dulR,t\right)=\mathbf P_{\nu}-{\bf A}_\nu$ (this second relation follow from Eq.~(7)). Therefore,
\begin{align}
 \nabla_\nu \gamma_l = \mathbf p_{l,\nu} - \left(\mathbf P_{\nu}-{\bf A}_\nu\right),
 \label{eqn: semiclassical relation between momenta}
\end{align}
with all quantities evaluated along the classical trajectory. When the (total) time derivative acts on both sides, it is evident that the left hand side, $\nabla_\nu\dot\gamma_l^{(I)}(t)$, is the difference between the classical forces $\dot{\mathbf p}_{l,\nu}$ and $\dot{\mathbf P}_{\nu}-\dot{\mathbf A}_{\nu} =0$. This second term is zero by virtue of Eq.~(8) for the classical force where the gauge condition~(9) is imposed. The first term instead can be interpreted as the force driving the (semiclassical) evolution of the BO-projected $l$-th wave packet, i.e. $-\nabla_\nu\epsilon_{BO}^{(l)}$, which yields $\nabla_\nu\dot\gamma_l^{(I)}(t)$ with no further contribution. This proves that Eq.~(16) is consistent with a semiclassical interpretation of the wave packets dynamics.

\section{SI.3 - Quantum momentum for a multi-level system}
In a multi-level system, $\nabla_\nu |\chi|^2/|\chi|^2$ can be written as a sum of two-state contributions as
\begin{align}
 \frac{\nabla_\nu \left|\chi(\dulR,t)\right|^2}{2\left|\chi(\dulR,t)\right|^2} = 
 \frac{1}{2(N_{st}-1)}\sum_{(p,q)}^{N_{st}}&\Bigg[
 \frac{\nabla_{\nu}\left|F_p(\dulR,t)\right|^2+\nabla_{\nu}\left|F_q(\dulR,t)\right|^2}{\left|F_p(\dulR,t)\right|^2+\left|F_q(\dulR,t)\right|^2}\left(|C_p(\dulR,t)|^2+|C_q(\dulR,t)|^2\right)\Bigg],
 \label{eqn: nabla chi/chi as sum of pairs}
\end{align}
where the sum over the indices $(p,q)$ involves pair of indexes with $p\neq q$ up to the total number of adiabatic states $N_{st}$ included in the calculation. The pre-factor $(N_{st}-1)^{-1}$ is introduced to avoid double counting. The term in the electronic equation~(29) containing the quantum momentum can then be re-written as
\begin{align}
\label{eqn: decoherence term multi level}
\boldsymbol{\mathcal P}_\nu^{(I)}\cdot\left(\tilde\bA^{(I)}_\nu-{\mathbf f}^{(I)}_{l,\nu}\right)C^{(I)}_l = \frac{-\hbar}{N_{st}-1}
\sum_{(p,q)}^{N_{st}}& \frac{\nabla_\nu \chi_{pq}^{(I)}(t)}{\chi_{pq}^{(I)}(t)}
\left(\rho^{(I)}_{pp}(t)+\rho^{(I)}_{qq}(t)\right)\nonumber\\
&\left(\sum_{k=1}^{N_{st}} \rho^{(I)}_{kk}(t) {\mathbf f}^{(I)}_{k,\nu}(t) - \sum_{k=1}^{N_{st}} \rho^{(I)}_{kk}(t) {\mathbf f}^{(I)}_{l,\nu}(t)\right)C^{(I)}_l  
\end{align}
where we have introduced the symbols $\chi_{pq}^{(I)}(t)=\sqrt{|F_p^{(I)}(t)|^2+|F_q^{(I)}(t)|^2}$ and $\tilde\bA^{(I)}_\nu$
\begin{align}
\tilde\bA^{(I)}_\nu(t) = \sum_{l=1}^{N_{st}}\rho_{ll}^{(I)}(t)\mathbf f_{l,\nu}^{(I)}(t),
\end{align}
meaning that, as also done in Eqs.~(28) and~(29), we neglect terms presenting the products of the quantum momentum and the non-adiabatic coupling vectors (NACVs)  since they are expressions of two competing effects, decoherence and population transfer. In the last line of Eq.~(\ref{eqn: decoherence term multi level}) we have used the PNC to write
\begin{align}
{\mathbf f}^{(I)}_{l,\nu}(t)=\sum_{k=1}^{N_{st}} \rho^{(I)}_{kk}(t) {\mathbf f}^{(I)}_{l,\nu}(t).
\end{align}
Eq.~(\ref{eqn: decoherence term multi level}) contains multiple sums over the BO states, each term including multiple products of BO populations/coefficients, such as $\rho^{(I)}_{pp} \rho^{(I)}_{kk} C^{(I)}_l$. To simplify this expression we assume that only the terms containing two BO states survive. For example, in a three-state system, when the indexes of the first sum are $(p,q)=(1,2)$, then the second sum runs over $k=1,2$, when $(p,q)=(1,3)$ then $k=1,3$, and so on. Eq.~(\ref{eqn: decoherence term multi level}) thus yields
\begin{align}
\label{eqn: approximated decohrence term multi level}
\boldsymbol{\mathcal P}_\nu^{(I)}\cdot\left(\tilde\bA^{(I)}_\nu-{\mathbf f}^{(I)}_{l,\nu}\right)\simeq \frac{1}{N_{st}-1}
\sum_{k=1}^{N_{st}}
\frac{\nabla_{\nu}\chi^{(I)}_{lk}(t)}{\chi^{(I)}_{lk}(t)}
\left(\rho^{(I)}_{kk}(t)+\rho^{(I)}_{ll}(t)\right)\rho^{(I)}_{kk}(t)\left({\mathbf f}^{(I)}_{k,\nu}(t) - {\mathbf f}^{(I)}_{l,\nu}(t)\right).  
\end{align}
At this point, the calculation of $\nabla_\nu\chi^{(I)}_{lk}(t)/\chi^{(I)}_{lk}$ follows the same procedure as for the two-level case, namely $\nabla_\nu\chi^{(I)}_{lk}(t)/\chi^{(I)}_{lk}(t)\propto\alpha_{lk,\nu}\left(\mathbf R_\nu^{(I)}(t) - \mathbf R_{lk,\nu}^0(t)\right)$, where the point $\mathbf R_{lk,\nu}^0$ is obtained as
\begin{align}
\label{eqn: crossover point for multi state}
R^0_{lk,\nu i} = \frac{\sum_I^{N_{traj}} R_{\nu i}^{(I)}(\rho^{(I)}_{ll}+\rho^{(I)}_{kk})\rho^{(I)}_{ll} 
\rho^{(I)}_{kk} (f^{(I)}_{k,\nu i}-f^{(I)}_{l,\nu 
i})}{\sum_I^{N_{traj}}(\rho^{(I)}_{ll}+\rho^{(I)}_{kk})\rho^{(I)}_{ll}\rho^{(I)}_{kk} (f^{(I)}_{k,\nu i}-f^{(I)}_{l,\nu i})}.
\end{align}
A similar procedure is employed to determine the expression of the last term in Eq.~(27).

\section{SI.4 - Simulations details}
The model systems employed for the numerical tests presented in the paper have been chosen as prototypes of non-adiabatic processes and may represent problematic situations for other trajectory-based quantum-classical approaches. Therefore, we aim at testing the performance of the CT-MQC algorithm developed in the paper against a quantum wave packet propagation scheme and other approximate schemes. We select the Ehrenfest approach~\cite{tully_fardisc1998,ehrenfest}, the trajectory surface hopping (TSH) approach in the fewest-switches version~\cite{tully1990} and independent-trajectory version of the MQC algorithm proposed here and based on the exact factorization~\cite{Gross_EPL2014, Gross_JCP2014}.

The Hamiltonian describing the two-state one-dimensional model systems is
\begin{align}
\hat H =\frac{-\hbar^2\partial_R^2}{2M} +\hat H_{d}(R)
\end{align}
where the electronic Hamiltonian $\hat H_{d}(R)$ in the diabatic basis is
\begin{align}
\hat H_{d}(R) = \left(
\begin{array}{cc}
H_{11}(R) & H_{12}(R) \\
H_{12}(R) & H_{22}(R)
\end{array}
\right).
\end{align}
The elements of this Hamiltonian will be defined below.
In all problems the nuclear mass is chosen to be $M=2000$, the proton mass in atomic units.

Quantum results are obtained from wave packets dynamics calculations using the split-operator~\cite{spo} technique. A time-step of  0.1~a.u. is used for the quantum simulations. The initial state is chosen such that only the lower BO state is populated and the nuclear wave packet is a normalized Gaussian centered at $R=-8$~a.u., model (a) and (b), at $R=-15$~a.u., model (c) and (d). The variances of the Gaussian wave packets at time 0 are chosen to be 20 times the inverse of the initial momenta.
%When we study the dependence of the final populations of the adiabatic states on the initial momentum for model (d), the initial variance is chosen to be 100 times the inverse of the initial momenta~\cite{tavernelliJCP2013}.
The results of the quantum-classical methods, MQC, Ehrenfest dynamics, TSH and CT-MQC, are obtained by integrating the electronic and nuclear equations of motion with time-step $0.5$~a.u., using the fourth-order Runge-Kutta and the velocity-Verlet algorithms, respectively. The initial conditions, positions and momenta, for the 200 trajectories used in CT-MQC, MQC and Ehrenfest and for the 5000 trajectories used in TSH, are sampled from the Wigner distribution corresponding to the initial quantum nuclear density. 

The populations of the BO states at time $t$ calculated in the approximate schemes are compared with the quantity
\begin{align}\label{eqn: BO populations}
\rho_{ll}(t) = \int d\dulR\,\left|F_l(\dulR,t)\right|^2,
\end{align}
obtained from wave packets dynamics. In the CT-MQC, MQC and Ehrenfest, the populations of the adiabatic states are averaged over their values for each trajectory, whereas in TSH the quantity $N_l(t)/N_{traj}$ indicates the population of the state $l$ ($N_l(t)$ is the number of trajectories running on the $l$-th BO surface at time $t$).

The indicator used to give a quantitative estimate of electronic decoherence is
\begin{align}\label{eqn: coherence mqc}
\left|\rho_{12}(t)\right|^2 = \frac{1}{N_{traj}}\sum_{I=1}^{N_{traj}} \left|C_1^{(I)}(t)\right|^2\left|C_2^{(I)}(t)\right|^2,
\end{align}
whose corresponding quantum mechanical counterpart is
\begin{align}\label{eqn: coherence}
 \left|\rho_{12}(t)\right|^2 = \int d\dulR\,\left|C_1\left(\dulR,t\right)\right|^2\left|C_2\left(\dulR,t\right)\right|^2
\left|\chi\left(\dulR,t\right)\right|^2.
\end{align}
Eq.~(\ref{eqn: coherence mqc}) is obtained from Eq.~(\ref{eqn: coherence}) using the following approximation for the nuclear density
\begin{align}
 \left|\chi\left(\dulR,t\right)\right|^2 = \frac{1}{N_{traj}}\sum_{I=1}^{N_{traj}} \delta\left(\dulR-\dulR^{(I)}(t)\right),
\end{align}
expression used to construct the histogram from the distribution of classical trajectories.

The problems studied in the paper~\cite{tully1990,tavernelliJCP2013} are (a) single avoided crossing, (b) dual avoided crossing, (c) extended coupling region with reflection and (d) double arch.

\subsection{Single avoided crossing}
The elements of the electronic Hamiltonian in the diabatic basis for model (a) are
\begin{equation}
\begin{array}{lll}
H_{11}(R) &=& a\left[1-\exp{\left(-bR\right)}\right], \,\, R>0, \\
H_{11}(R) &=& -a\left[1-\exp{\left(bR\right)}\right], \,\, R<0, \\
H_{22}(R) &=& -H_{11}(R), \\
H_{12}(R) &=& H_{21}(R) = c\exp{\left(-dR^2\right)},
\end{array}
\end{equation}
with $a=0.01,\,b=1.6\,,c=0.005,\,d=1.0$. 

\subsection{Dual avoided crossing}
The elements of the electronic Hamiltonian in the diabatic basis for model (b) are
\begin{equation}
\begin{array}{lll}
H_{11}(R) &=& 0, \\
H_{22}(R) &=& -a\exp{\left(-bR^2\right)}+e_0, \\
H_{12}(R) &=& H_{21}(R) = c\exp{\left(-dR^2\right)},
\end{array}
\end{equation}
with $a=0.1,\,b=0.28\,,c=0.015,\,d=0.06,\,e_0=0.05$.

\subsection{Extended coupling region with reflection}
The elements of the electronic Hamiltonian in the diabatic basis for model (c) are
\begin{equation}
\begin{array}{lll}
H_{11}(R) &=& a, H_{22}(R) = -a,\\
H_{12}(R) &=& b\exp{\left(cR\right)}, R<0,\\
H_{12}(R) &=& b\left[2-\exp{\left(-cR\right)}\right], R>0,\\
H_{21}(R) &=& H_{12}(R) 
\end{array}
\end{equation}
with $a=6\times 10^{-4},\,b=0.1,\,c=0.9$.

\subsection{Double arch}
The elements of the electronic Hamiltonian in the diabatic basis for model (d) are
\begin{equation}
\begin{array}{lll}
H_{11}(R) &=& a, H_{22}(R) = -a,\\
H_{12}(R) &=& -b\exp{(c(R-d))}\\
                    &&+b\exp{(c(R+d))}, R<-d,\\
H_{12}(R) &=& b\exp{(-c(R-d))}\\
                   &&-b\exp{(-c(R+d))}, R<d,\\
H_{12}(R) &=& 2b-b\exp{(c(R-d))}\\
                   && -b\exp{(-c(R+d))}, -d<R<d,\\
H_{21}(R) &=& H_{12}(R) 
\end{array}
\end{equation}
with $a=6\times 10^{-4},\,b=0.1,\,c=0.9,\,d=4$.

\addcontentsline{toc}{section}{References}
%\bibliography{./biblio,./factorization,./mqc}